\documentclass{article}

\usepackage{graphicx}
\usepackage{booktabs}
\usepackage{array}
\usepackage{geometry}
\usepackage{siunitx}
\usepackage{multirow}
\usepackage{caption}
\usepackage{subcaption}
\usepackage{float}
\usepackage{threeparttable}
\usepackage{booktabs}
\usepackage{multirow}
\usepackage{graphicx}
\usepackage{comment}
\usepackage{siunitx}
\usepackage{cite}   

\usepackage{threeparttable}
\title{A GRACE Interatomic Potential for W--Ta Systems: Application to Ta Segregation and Pinning of Shear-Coupled Grain-Boundary Migration}

\author{%
Omar Hussein$^{1}$,
Laurent Capolungo$^{1}$, 
Danny Perez$^{2}$\\
\small $^{1}$Materials Science and Technology Division, MST-8, Los Alamos National Laboratory,
Los Alamos, NM, 87545, USA\\
\small $^{2}$X-Computational Physics Division, XCP-AI4ND, Los Alamos National Laboratory,
Los Alamos, NM, 87545, USA\\
}

\date{}
\begin{document}

\maketitle
\section{Abstract}

Tungsten–tantalum alloys are promising plasma-facing materials for fusion reactors, but simulations of their coupled defect, mechanical, and microstructural evolution require an interatomic potential that remains accurate across diverse atomic environments. We develop a transferable W–Ta potential within the GRaph Atomic Cluster Expansion (GRACE) framework. The primary training dataset was generated automatically by maximizing information entropy in an atomically relevant feature space, avoiding manual enumeration of the structures and processes required for specific downstream applications.  Extensive validation demonstrates a consistent description of unary and binary phase stability, elasticity, phonons, high-pressure response, surfaces, point defects, dislocation energetics, transformation pathways, alloy thermodynamics, and melting behavior.

As a demonstration, the potential is applied to Ta segregation and shear-coupled migration of a $\Sigma5(120)$ W GB. Simulations reveal strongly site-selective segregation: Ta preferentially occupies a small subset of crystallographically distinct atomic rows. Under shear, segregated Ta impedes disconnection-mediated grain boundary migration. The de-pinning stress initially increases with Ta concentration but plateaus as the dominant pinning environment becomes saturated, while increasing temperature weakens both segregation and pinning. These results establish a broadly validated W–Ta potential and demonstrate how automated entropy-maximized data generation can produce transferable atomistic models capable of resolving mechanisms not explicitly targeted during training.

\section{Introduction}

Tungsten (W) and W-based alloys are leading candidate materials for plasma-facing components in fusion reactors because of their high melting temperature, high thermal conductivity, low sputtering yield, and low hydrogen-isotope retention. Despite these advantages, the use of W is limited by low-temperature brittleness, poor workability, recrystallization-induced embrittlement, and neutron-irradiation-induced degradation of mechanical properties~\cite{philipps2011tungsten,neu2005tungsten,rieth2013recent,coenen2017advanced,nogami2022tungsten,gumbsch2003brittle,giannattasio2007strain}. These limitations have motivated several alloying and microstructural design strategies, including grain refinement~\cite{farrell1967recrystallization}, work hardening~\cite{wei2008effect}, dispersion strengthening~\cite{schade2010100}, K doping~\cite{nogami2019improvement}, Re alloying~\cite{geach1955alloys,romaner2010effect}, and solid-solution alloying~\cite{luo1991solution}.

Tantalum (Ta) is a particularly relevant alloying element for W because W and Ta form a high-temperature solid solution and Ta has been considered more favorable than some refractory solutes from the standpoint of fusion-relevant activation and transmutation behavior~\cite{makarov2002development,noda1988materials,maziasz1990precipitation}. Recent experimental studies have shown that small Ta additions, typically in the range of 1--5 wt.\%, can improve strength, recrystallization resistance, irradiation tolerance, and high-heat-flux performance relative to pure W~\cite{nogami2022tungsten,nogami2021thermal,miyazawa2023microstructural,du2023properties}. Ta also modifies plasma-induced surface evolution, including D/He retention, blistering, and fuzz formation~\cite{schmid2012comparison,zayachuk2013surface,li2020effect}. On the other hand, studies of W--Ta alloys reported limited or even detrimental effects on ductility and fracture toughness, indicating a strong dependence on processing history, impurity content, grain and subgrain structure, and the spatial distribution of Ta~\cite{wurster2011fracture,rieth2012impact}. Therefore, the role of Ta cannot be explained solely by bulk solid-solution strengthening; its interactions with defects, interfaces, and grain boundaries (GBs) must also be considered.

The W--Ta system remains challenging even from the thermodynamic standpoint. Although the experimental phase diagram suggests high-temperature miscibility without stable intermetallic compounds, first-principles and cluster-expansion studies have predicted competing low-temperature ordered phases~\cite{maziasz1990precipitation,muzyk2011phase,turchi2001first,hart2005evolutionary}. 

This thermodynamic complexity is compounded by the absence of a transferable W--Ta interatomic potential capable of describing the coupled kinetic processes that govern fusion-material performance. In service, W-based alloys simultaneously undergo irradiation damage, defect migration, solute redistribution, recrystallization, plastic flow, high-temperature deformation, and plasma--surface evolution. Modeling these processes requires accuracy far beyond equilibrium bulk properties: the potential must remain reliable across compositions, defects, interfaces, transformation pathways, and extreme thermomechanical conditions.

Existing Embedded Atom Method/Finnis--Sinclair (EAM/FS) potentials for W--Ta~\cite{chen2019development} are limited in their ability to represent angular bonding, higher-order correlations, body-centered cubic (BCC) defect physics, alloy ordering, interface structures, and migration barriers. More recent Generalized EAM (GEAM) efforts~\cite{sharma2023development} introduced higher-order interactions, but were trained on small datasets and validated over a limited set of properties, with restricted assessment of dislocation energetics, generalized stacking-fault energies, point-defect migration, binary defect interactions, solute clustering, and GB response. These limitations motivate the development of a transferable machine-learning interatomic potential for W--Ta that can be applied to the complex, coupled phenomena expected in extreme fusion environments.

In this work, we develop and validate an ultra-robust GRaph Atomic Cluster Expansion (GRACE) potential for the W--Ta binary system~\cite{drautz2019atomic,bochkarev2024graph,pa2025information}. The potential  describes unary and binary energetics, elastic properties, defect structures, transformation pathways, interfacial environments, and high-temperature configurations relevant to W--Ta alloys. The central goal is to provide a robust atomistic model that can be used to study fusion-relevant processes in W--Ta, including irradiation defect evolution, recrystallization, plastic deformation, solute effects, and GB-mediated deformation. As an application, we use this potential to investigate how Ta segregates at high--angle tilt GBs and affects shear-driven GB motion in W, with particular focus on the dependence of \emph{GB de-pinning} on the segregated Ta atmosphere. By connecting solute content, GB structure, and atomistic de-pinning mechanisms, this study investigates how binary alloying controls GB-mediated deformation in W-based refractory alloys while establishing a transferable potential for broader simulations of W--Ta in extreme environments.

\section{Training Data Generation}

The majority of the training data was generated using the feature information entropy maximization approach~\cite{karabin2020entropy,pa2025information,bochkarev2026exploring}. At a high level, the method reframes the problem of generating training sets for MLIPs as an optimization under an objective function that quantifies the diversity of the dataset as measured in an atomically-relevant feature space. 
Using a Gaussian upper bound to the Shannon differential entropy of the feature distribution, the procedure optimizes the information content of the entire dataset by displacing atoms along the gradient of the information entropy.  This greedy optimization procedure is completely automated and leads to ultra-diverse dataset, which in turn leads to extremely transferable models across a broad range of properties and conditions~\cite{montes2022training,pa2025information,bochkarev2026exploring}.

\begin{figure}[t]
\centering
\includegraphics[width=0.8\textwidth]{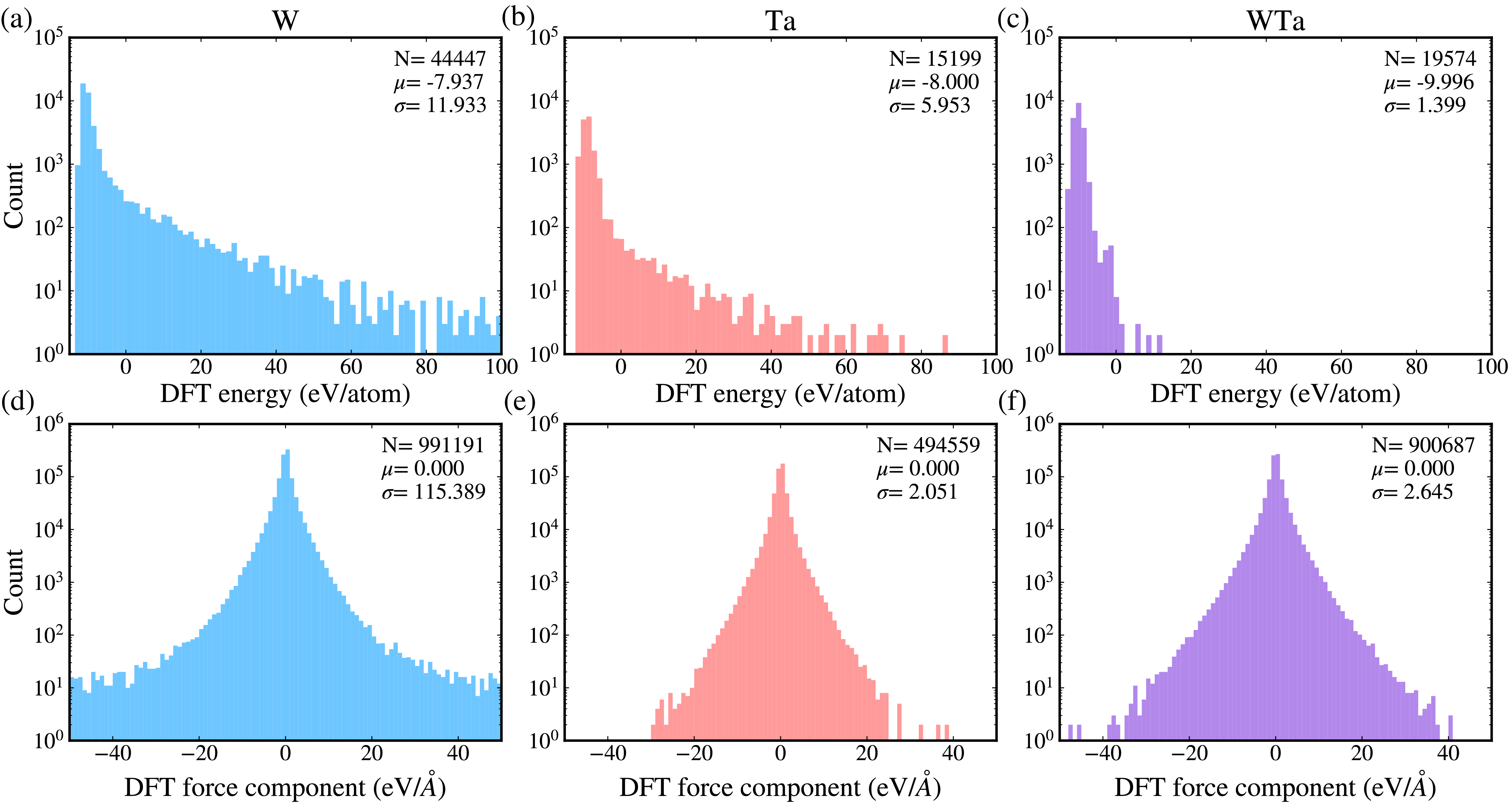}
\caption{
Distributions of DFT energies (a--c) and Cartesian force components (d--f) in the W--Ta training database. The columns correspond to (a,d) W, (b,e) Ta, and (c,f) mixed W--Ta configurations. Counts are plotted on a logarithmic scale, and the annotations report the number of samples $N$, mean $\mu$, and standard deviation $\sigma$ for each distribution.
}
\label{fig:Dataset_Distribution}
\end{figure}

In this work, we follow the data-generation and DFT-evaluation protocol of Ref.~\cite{bochkarev2026exploring}, restricting the chemical space to W and Ta. Using this procedure, we generated 78,703 configurations containing 788,758 atoms. The dataset was further enriched with 517 intermetallic configurations from the OQMD database~\cite{saal2013materials} and previous DFT studies~\cite{muzyk2011phase,blum2005prediction}. The distributions of DFT energies and forces used for model training are shown in Fig.~\ref{fig:Dataset_Distribution}.

Density functional theory (DFT) calculations were performed using the Vienna Ab initio Simulation Package (VASP)~\cite{kresse1996} through the Atomic Simulation Environment (ASE)~\cite{hjorth2017atomic}. The calculations employed the projector-augmented-wave method and the Perdew–Burke–Ernzerhof (PBE) generalized-gradient approximation for the exchange–correlation functional~\cite{perdew1996generalized}. 

Single-point calculations were conducted on the supplied atomic configurations without ionic-position or simulation-cell relaxation. Periodic boundary conditions were applied in all three spatial directions. The plane-wave kinetic-energy cutoff was set to ($500~\mathrm{eV}$), and Brillouin-zone sampling was generated automatically using a reciprocal-space spacing of ($0.125~\mathrm{\AA^{-1}}$). Gaussian smearing was applied with a smearing width of ($0.1~\mathrm{eV}$). All calculations were performed without spin polarization. The electronic self-consistent-field iterations were converged to an energy tolerance of ($10^{-6}~\mathrm{eV}$), with a maximum of 200 electronic iterations. 

\section{GRACE Model Training}
The W--Ta interatomic potential was developed using the Graph Atomic Cluster Expansion (GRACE) framework~\cite{bochkarev2024graph}. A one-layer GRACE model was used with a $6.0~\mathrm{\AA}$ cutoff, maximum body order of four, angular resolution $l_{\max}=4$, $n_{\mathrm{rad}}^{\max}=32$ radial channels, $n_{\mathrm{rad}}^{\mathrm{base}}=8$ radial basis functions, $n_{\mathrm{prod}}^{\max}=32$ product functions, and $n_{\mathrm{MLP}}^{\mathrm{dens}}=10$ density-embedding hidden units. A short-range ZBL correction was included with a cutoff of $1.8~\mathrm{\AA}$.

The W--Ta convex hull was used to define an energy-based weighting scheme, where configurations closer to the hull were assigned larger statistical weights to emphasize low-energy crystalline, alloy, and defect structures. Higher-energy configurations were retained with reduced weights to improve transferability to distorted and non-equilibrium environments. The dataset was split into training and test sets using a 5\% test fraction. Model parameters were optimized for 1500 epochs using the Adam optimizer with an initial learning rate of 0.008, cosine decay to $6.67\times10^{-4}$, exponential moving averaging with momentum 0.99, and gradient clipping with a maximum norm of 1.0. Huber losses with $\delta=0.01$ were used for both energies and forces, with relative loss weights of 10 and 5, respectively.

\begin{figure}[t]
\centering
\includegraphics[width=0.8\textwidth]{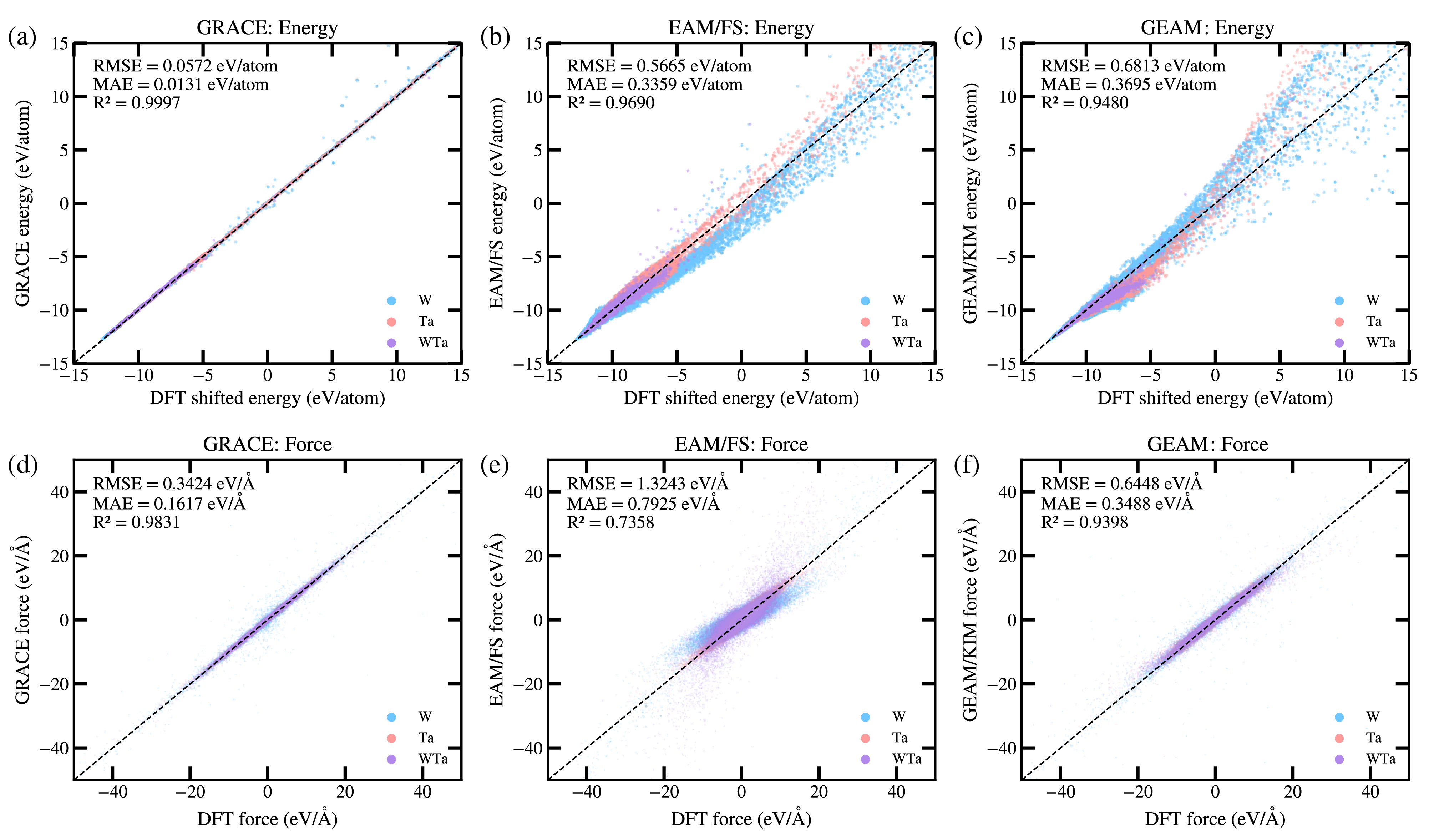}
\caption{
Parity plots comparing DFT energies and atomic force components with predictions from GRACE, EAM/FS, and GEAM for W, Ta, and W--Ta configurations. Panels (a)--(c) show shifted energies per atom, and panels (d)--(f) show force components. Dashed lines indicate perfect agreement with DFT.
}
\label{fig:parity}
\end{figure}

The predictive performance of GRACE on the training and testing datasets is compared with the available EAM/FS~\cite{chen2019development} and GEAM~\cite{sharma2023development} W--Ta potentials in Fig.~\ref{fig:parity} and Table~\ref{tab:parity_metrics}. It is important to note that the EAM/FS and GEAM potentials were not parameterized using the present  DFT dataset. Therefore, this comparison is not intended as a direct, like-for-like assessment of fitting accuracy, but rather as an evaluation of how robustly the different potentials reproduce an ultra-diverse set of W--Ta configurations that extensively probes different regions of the energy landscape.  GRACE gives the lowest energy errors, with $\mathrm{RMSE}=0.0572~\mathrm{eV/atom}$ and $\mathrm{MAE}=0.0131~\mathrm{eV/atom}$, compared with $0.5665$ and $0.3359~\mathrm{eV/atom}$ for EAM/FS and $0.6813$ and $0.3695~\mathrm{eV/atom}$ for GEAM, respectively. GRACE also provides the most accurate forces, with $\mathrm{RMSE}=0.3424~\mathrm{eV/\AA}$, $\mathrm{MAE}=0.1617~\mathrm{eV/\AA}$, and $R^2=0.9831$. 

\begin{table}[htbp]
\centering
\caption{Comparison of prediction accuracy on the dataset. RMSE, MAE, and $R^2$ are reported for energies and atomic forces relative to DFT reference values.}
\label{tab:parity_metrics}
\begin{tabular}{lcccccc}
\hline
\textbf{Potential} &
\textbf{E-RMSE} &
\textbf{E-MAE} &
\textbf{E-$R^2$} &
\textbf{F-RMSE} &
\textbf{F-MAE} &
\textbf{F-$R^2$} \\
&
\textbf{(eV/atom)} &
\textbf{(eV/atom)} &
&
\textbf{(eV/\AA)} &
\textbf{(eV/\AA)} &
\\
\hline
GRACE    & 0.0572 & 0.0131 & 0.9997 & 0.3424 & 0.1617 & 0.9831 \\
EAM/FS   & 0.5665 & 0.3359 & 0.9690 & 1.3243 & 0.7925 & 0.7358 \\
GEAM & 0.6813 & 0.3695 & 0.9480 & 0.6448 & 0.3488 & 0.9398 \\
\hline
\end{tabular}
\end{table}

\section{GRACE Model Validation}

The W--Ta GRACE machine-learning interatomic potential was validated against a broad set of unary and binary properties to assess its accuracy and transferability. All GRACE atomistic simulations were performed using the Large-scale Atomic/Molecular Massively Parallel Simulator (LAMMPS)~\cite{LAMMPS}. 

\subsection{Unary Properties Validation}

Table~\ref{tab:w_ta_unary_properties} shows that the GRACE potential reproduces a range of unary properties of BCC W and Ta with good overall agreement relative to the present DFT calculations and available experimental data. For both elements, the predicted equilibrium lattice constants, cohesive energies, bulk moduli, and elastic constants are close to the DFT values, indicating that the model accurately captures the near-equilibrium BCC response. The vacancy formation and migration energies are also well reproduced, while the interstitial formation energies preserve the correct energetic ordering among the different defect configurations. In addition, the surface and twin-boundary energies predicted by GRACE remain close to the DFT references, demonstrating that the potential describes low-coordination and interfacial environments with reasonable accuracy.

\begin{table}[H]
\centering
\caption{Unary properties of BCC W and Ta comparing experiment, present DFT, and GRACE predictions.}
\label{tab:w_ta_unary_properties}
\resizebox{\textwidth}{!}{
\begin{tabular}{lcccccc}
\toprule
\multirow{2}{*}{Property} 
& \multicolumn{3}{c}{W} 
& \multicolumn{3}{c}{Ta} \\
\cmidrule(lr){2-4} \cmidrule(lr){5-7}
& Experiment & DFT & GRACE 
& Experiment & DFT & GRACE \\
\midrule

$a_0$ (\si{\angstrom}) 
& 3.165\textsuperscript{a} & 3.172 & 3.184
& 3.3039\textsuperscript{b} & 3.309 & 3.322 \\

$E_0$ (\si{\electronvolt}) 
& -8.9\textsuperscript{c} & -13.018 & -12.956
& -8.1\textsuperscript{d} & -11.865 & -11.815 \\

$B$ (\si{\giga\pascal}) 
& 308--314\textsuperscript{e} & 320.728 & 326.089
& 194.2\textsuperscript{f}; 198\textsuperscript{b} & 196.537 & 191.164 \\

$C_{11}$ (\si{\giga\pascal}) 
& 501--521\textsuperscript{e} & 551.001 & 561.991
& 266.3\textsuperscript{f} & 264.516 & 252.070 \\

$C_{12}$ (\si{\giga\pascal}) 
& 199--207\textsuperscript{e} & 205.591 & 208.138
& 158.2\textsuperscript{f} & 160.565 & 160.711 \\

$C_{44}$ (\si{\giga\pascal}) 
& 151--160\textsuperscript{e} & 148.866 & 153.278
& 87.4\textsuperscript{f} & 74.097 & 69.199 \\

$E_v^f$ (\si{\electronvolt}) 
& 3.1--4.0\textsuperscript{g} & 3.223 & 3.480
& 2.2--3.1\textsuperscript{h} & 3.018 & 3.011 \\

$E_v^m$ (\si{\electronvolt}) 
& 1.7--2.0\textsuperscript{i} & 1.683 & 1.842
& 0.7\textsuperscript{h} & 0.656 & 0.640 \\

$E_i^f \langle 100 \rangle$ (\si{\electronvolt}) 
& -- & 11.985 & 11.332
& -- & 5.699 & 6.586 \\

$E_i^f \langle 110 \rangle$ (\si{\electronvolt}) 
& -- & 10.176 & 10.845
& -- & 5.213 & 6.041 \\

$E_i^f \langle 111 \rangle$ (\si{\electronvolt}) 
& -- & 9.824 & 10.432
& -- & 4.572 & 5.174 \\

$E_i^f$ Tet. (\si{\electronvolt}) 
& -- & 11.393 & 11.406
& -- & 5.577 & 6.602 \\

$E_i^f$ Oct. (\si{\electronvolt}) 
& -- & 12.063 & 11.660
& -- & 5.776 & 6.824 \\

$\gamma^s_{100}$ (\si{\joule\per\meter\squared}) 
& -- & 4.011 & 3.827
& -- & 2.503 & 2.637 \\

$\gamma^s_{110}$ (\si{\joule\per\meter\squared}) 
& -- & 3.252 & 3.298
& -- & 2.372 & 2.409 \\

$\gamma^s_{111}$ (\si{\joule\per\meter\squared}) 
& -- & 3.496 & 3.565
& -- & 2.720 & 2.768 \\

$\gamma^s_{211}$ (\si{\joule\per\meter\squared}) 
& -- & 3.988 & 3.808
& -- & 3.025 & 2.863 \\

$\gamma^\mathrm{TB}_{211}$ (\si{\joule\per\meter\squared}) 
& -- & 0.760\textsuperscript{j} & 0.840
& -- & 0.304\textsuperscript{j} & 0.329 \\

$T_m$ (K) 
& 3695\textsuperscript{k} & 3450.000 $\pm$ 105.000\textsuperscript{l} & 3555.250 $\pm$ 40.270
& 3293\textsuperscript{c} & 3085.000 $\pm$ 130.000\textsuperscript{m} & 3051.140 $\pm$ 42.060 \\

\bottomrule
\end{tabular}
}
\begin{flushleft}
\footnotesize{
\textsuperscript{a}\cite{lide1988crc};
\textsuperscript{b}\cite{dewaele2004refinement};
\textsuperscript{c}\cite{kittel1996introduction};
\textsuperscript{d}\cite{kittel2018introduction};
\textsuperscript{e}\cite{bolef1962elastic};
\textsuperscript{f}\cite{featherston1963elastic};
\textsuperscript{g}\cite{maier1979high};
\textsuperscript{h}\cite{ehrhart1991atomic};
\textsuperscript{i}\cite{balluffi1978vacancy};
\textsuperscript{j}\cite{li2023theoretical};
\textsuperscript{k}\cite{rudy1969compendium};
\textsuperscript{l}\cite{wang2011melting};
\textsuperscript{m}\cite{taioli2007melting}.
\\
Values reported without a lettered literature footnote correspond to the present DFT or GRACE calculation.
}
\end{flushleft}
\end{table}

In the reference DFT calculations, bulk BCC cells were relaxed with respect to both atomic positions and cell dimensions, whereas defect and surface structures were relaxed at fixed cell dimensions. Elastic constants were obtained from energy--strain fits using hydrostatic, orthorhombic, and shear strains ranging from $\pm0.001$ to $\pm0.003$. Vacancy formation and migration energies were evaluated in $6\times6\times6$ supercells, with migration barriers determined using a climbing-image nudged elastic band (NEB)~\cite{jonsson1998nudged} calculation with 21 replicas, see subsection~\ref{subsubsec:vacancy_migration_barrier_cal}. Self-interstitial formation energies were calculated in $6\times6\times6$ BCC supercells for the $\langle100\rangle$, $\langle110\rangle$, and $\langle111\rangle$ dumbbells and the tetrahedral and octahedral configurations. Surface energies were calculated from relaxed slabs containing two equivalent free surfaces as $\gamma=(E_{\mathrm{slab}}-N E_{\mathrm{bulk}})/(2A)$.

For the GRACE property calculation, the BCC lattice constants and cohesive energies were obtained from periodic two-atom cells relaxed at zero pressure. Elastic constants were calculated from relaxed $8\times8\times8$ BCC supercells using central finite differences, $C_{ij}=\partial \sigma_i/\partial \varepsilon_j$, over strains from $\pm1\times10^{-4}$ to $\pm1\times10^{-3}$. The tensor was symmetrized and averaged, and the bulk modulus was evaluated as $B=(C_{11}+2C_{12})/3$.

Vacancy formation energies were obtained by removing a central atom from periodic $6\times6\times6$ supercells and relaxing the remaining atoms at fixed cell dimensions, with $E_{\mathrm{f}}^{\mathrm{vac}}=E_{\mathrm{vac}}-(N-1) E_{\mathrm{BCC}}$. Migration barriers were calculated using the NEB method~\cite{henkelman2000improved,henkelman2000climbing} as $E_{\mathrm{m}}^{\mathrm{vac}}=E_{\mathrm{saddle}}-E_{\mathrm{initial}}$. Self-interstitial formation energies were calculated for the $\langle100\rangle$, $\langle110\rangle$, and $\langle111\rangle$ dumbbells and the octahedral and tetrahedral configurations in periodic $6\times6\times6$ supercells. After full relaxation, the formation energy was evaluated as $E_{\mathrm{f}}^{\mathrm{int}}=E_{\mathrm{int}}-(N+1)E_{\mathrm{BCC}}$. For the point-defect calculations, $N$ denotes the number of atoms in the corresponding perfect supercell, and $E_{\mathrm{BCC}}$ is the bulk energy per atom calculated using the same method. 

Surface energies were obtained from slabs periodic in-plane and free normal to the surface using $\gamma=(E_{\mathrm{slab}}-N E_{\mathrm{BCC}})/(2A)$. The $\Sigma3\{112\}\langle110\rangle$ coherent twin-boundary energy was calculated from a fully periodic bicrystal containing two equivalent boundaries using the corresponding excess-energy expression.

Melting temperatures were estimated from solid--liquid coexistence simulations in periodic $6\times6\times50$ BCC cells. A liquid slab was generated within the equilibrated solid, and the system was evolved under near-zero-pressure NPH conditions until a steady coexistence temperature was reached.

\subsubsection{Relative Phase Stability}
\begin{figure}[H] 
    \centering 
    \includegraphics[width=0.8\textwidth]{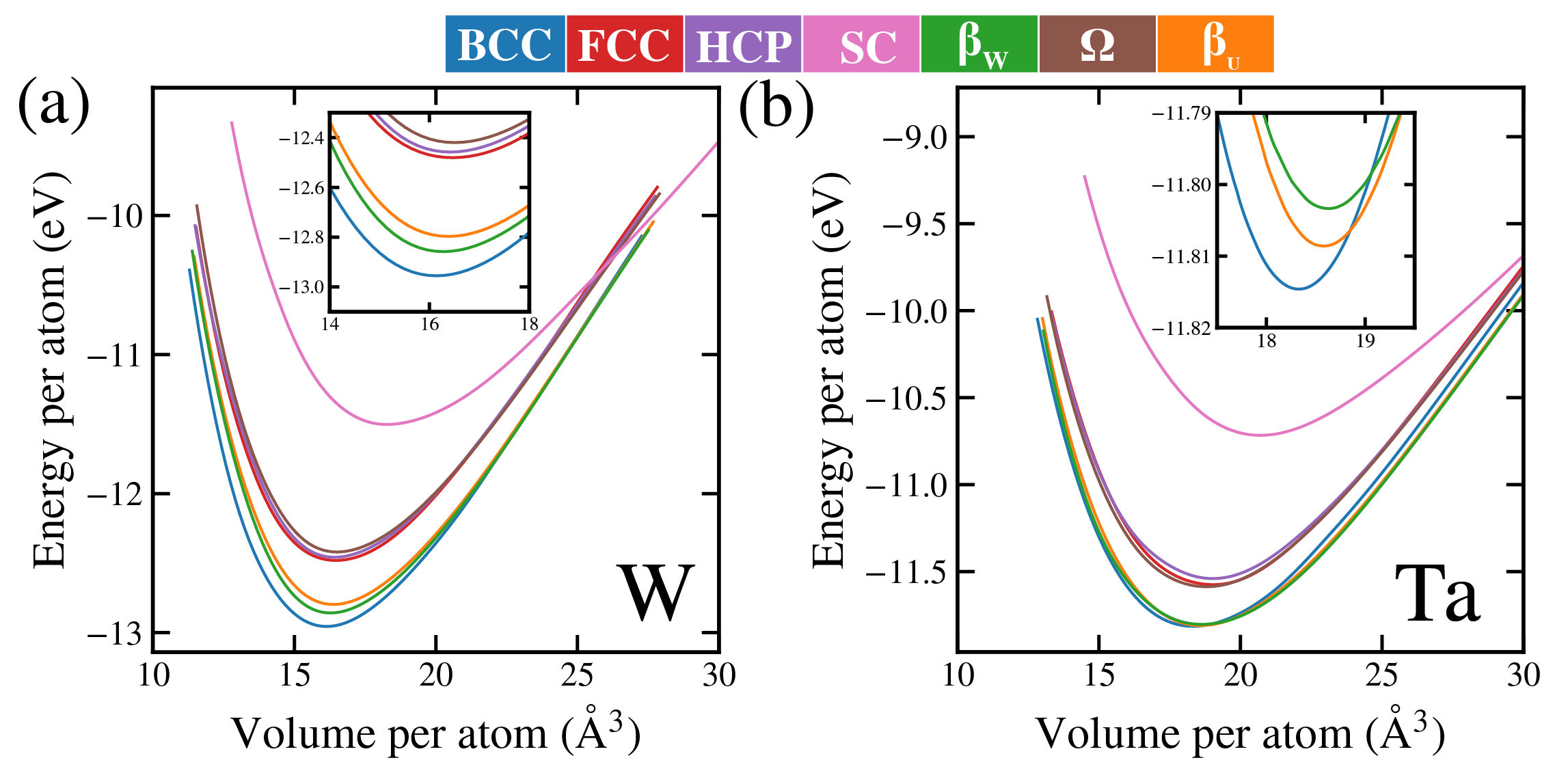} 
    \caption{Equation-of-state curves for competing crystal structures of pure (a) W and (b) Ta predicted by the GRACE potential.} 
\label{fig:eos_phase_stability} \end{figure}

The stability of pure W and Ta was assessed by comparing several crystal structures relative to the BCC phase. The equation-of-state curves were obtained by uniformly scaling the relaxed cell and evaluating the energy at each volume without further atomic relaxation. Phase energies were reported per atom relative to BCC as $\Delta E_{\mathrm{phase}}=E_{\mathrm{phase}}-E_{\mathrm{BCC}}$. As shown in Fig.~\ref{fig:eos_phase_stability} and Table~\ref{tab:w_ta_phase_stability}, GRACE correctly predicts BCC as the most stable structure for both elements. For W, GRACE closely matches the ab initio relative energies of FCC, SC, and A15, while the higher-energy DC, dimer, and trimer structures remain far above BCC. For Ta, GRACE also captures the main stability trend, including the low energy of A15 and the higher energies of FCC, HCP, HEX, SC, and DC. Some differences remain, especially for A15 and HCP Ta, but the overall ordering of the competing structures is well reproduced.

\begin{table}[H]
\centering
\caption{Relative phase stability for unary W and Ta. Energies are reported relative to the corresponding BCC ground state of each element. For dimer and trimer structures, $R_0$ is reported instead of $a_0$.}
\label{tab:w_ta_phase_stability}
\begin{threeparttable}
\resizebox{\textwidth}{!}{
\begin{tabular}{lcccccccc}
\toprule
\multirow{2}{*}{Structure}
& \multicolumn{2}{c}{Ab initio W}
& \multicolumn{2}{c}{GRACE W}
& \multicolumn{2}{c}{Ab initio Ta}
& \multicolumn{2}{c}{GRACE Ta} \\
\cmidrule(lr){2-3}
\cmidrule(lr){4-5}
\cmidrule(lr){6-7}
\cmidrule(lr){8-9}
& $\Delta E$ (\si{\electronvolt/atom})
& $a_0$ or $R_0$ (\si{\angstrom})
& $\Delta E$ (\si{\electronvolt/atom})
& $a_0$ or $R_0$ (\si{\angstrom})
& $\Delta E$ (\si{\electronvolt/atom})
& $a_0$ or $R_0$ (\si{\angstrom})
& $\Delta E$ (\si{\electronvolt/atom})
& $a_0$ or $R_0$ (\si{\angstrom}) \\
\midrule

FCC    
& 0.470\tnote{a} & 4.000\tnote{a} 
& 0.475 & 4.038 
& 0.304\tnote{c} & 4.1256\tnote{c} 
& 0.239 & 4.239 \\

HCP    
& -- & -- 
& 0.497 & 2.778 
& 0.385\tnote{c} & 2.9112\tnote{c} 
& 0.275 & 2.922 \\

HEX    
& -- & -- 
& 1.137 & 2.662 
& 0.757\tnote{c} & 2.7715\tnote{c} 
& 0.703 & 2.877 \\

SC     
& 1.350\tnote{a} & 2.610\tnote{a} 
& 1.452 & 2.634 
& 1.137\tnote{c} & 2.6710\tnote{c} 
& 1.098 & 2.746 \\

DC     
& 2.330\tnote{b} & 5.870\tnote{b} 
& 2.662 & 5.781 
& 3.128\tnote{c} & 5.9096\tnote{c} 
& 2.885 & 6.021 \\

$\beta$-W/A15   
& 0.080\tnote{a} & 5.060\tnote{a} 
& 0.097 & 5.069 
& 0.060\tnote{c} & 5.1805\tnote{c} 
& 0.011 & 5.301 \\

Dimer  
& -- & -- 
& 5.919 & 2.132 
& -- & 6.044\tnote{c} 
& 5.711 & 2.258 \\

Trimer 
& -- & -- 
& 4.849 & 2.286 
& -- & 5.944\tnote{c} 
& 4.574 & 2.417 \\

\bottomrule
\end{tabular}
}
\begin{tablenotes}
\small
\item[a] Ref.~\cite{mrovec2007bond}.
\item[b] Ref.~\cite{juslin2005analytical}.
\item[c] Ref.~\cite{mishin2006angular}.
\end{tablenotes}
\end{threeparttable}
\end{table}

\subsubsection{Thermal Expansion}
\begin{figure}[H]
\centering
\includegraphics[width=0.95\textwidth]{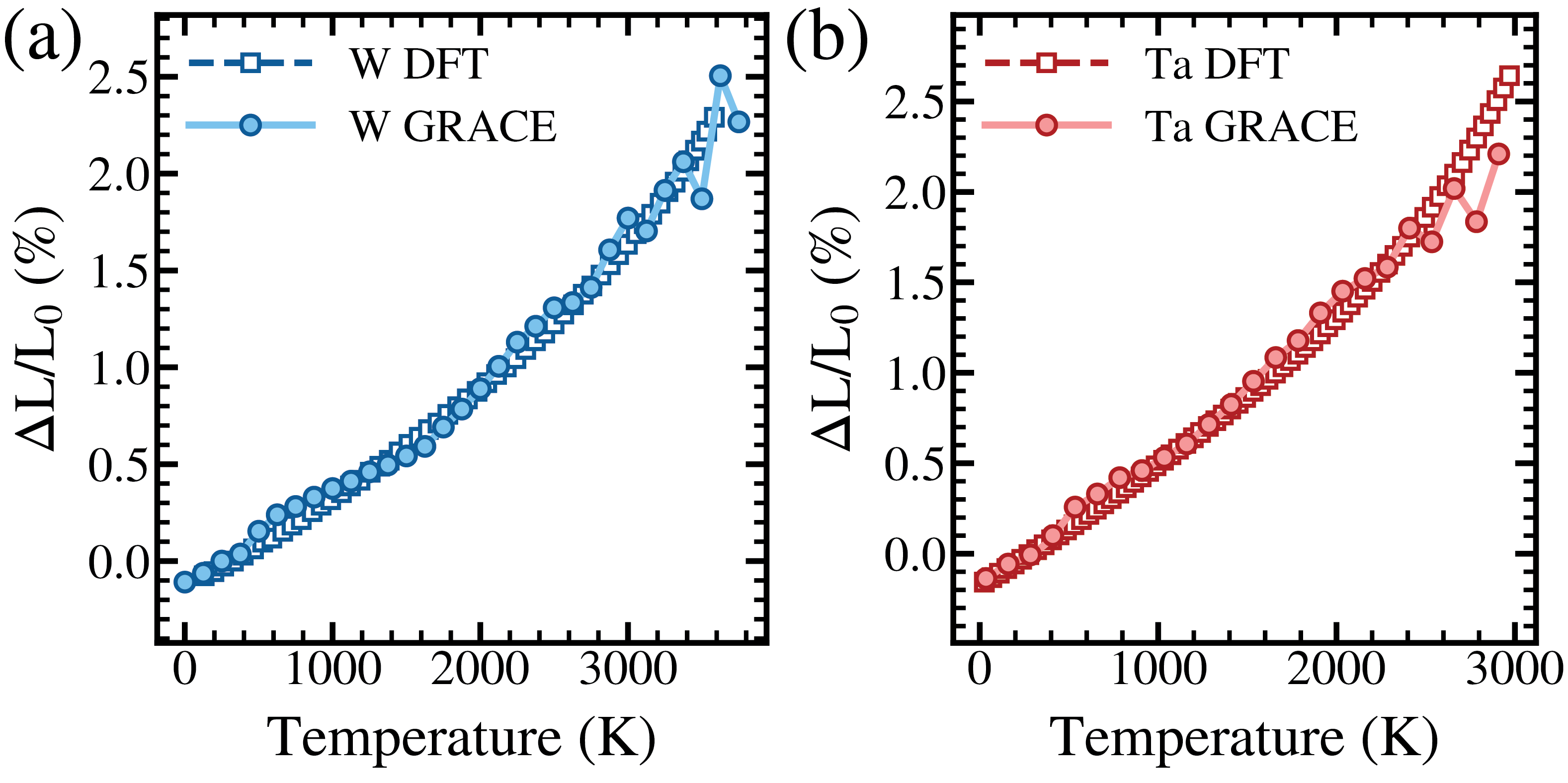}
\caption{Thermal expansion of BCC (a) W and (b) Ta, expressed as the relative change in lattice parameter, $\Delta L/L_0$. GRACE predictions are compared with DFT reference data for W~\cite{forslund2022ab} and Ta~\cite{ono2009first}.}
\label{fig:thermal-expansion}
\end{figure}

Thermal expansion was evaluated using periodic $6\times6\times6$ BCC supercells heated from $100$ to $3500~\mathrm{K}$ in $100~\mathrm{K}$ increments under zero-pressure NPT conditions. The relative expansion, $\Delta L/L_0=[L(T)-L_0]/L_0$, increased steadily with temperature for both W and Ta, as shown in Fig.~\ref{fig:thermal-expansion}. GRACE closely matches the DFT thermal expansion of W and reproduces the nonlinear DFT trend for Ta, with only minor high-temperature deviations.

\subsubsection{High Pressure}
High-pressure behavior was evaluated using periodic BCC supercells equilibrated near $300~\mathrm{K}$ and subsequently compressed isotropically in successive $0.5\%$ linear increments under NVT conditions. The pressure was averaged at each volume and plotted against $V/V_0$. GRACE accurately reproduces the strong nonlinear pressure increase upon compression for both W and Ta up to pressures in the hundreds of GPa.

\begin{figure}[H]
\centering
\includegraphics[width=0.95\textwidth]{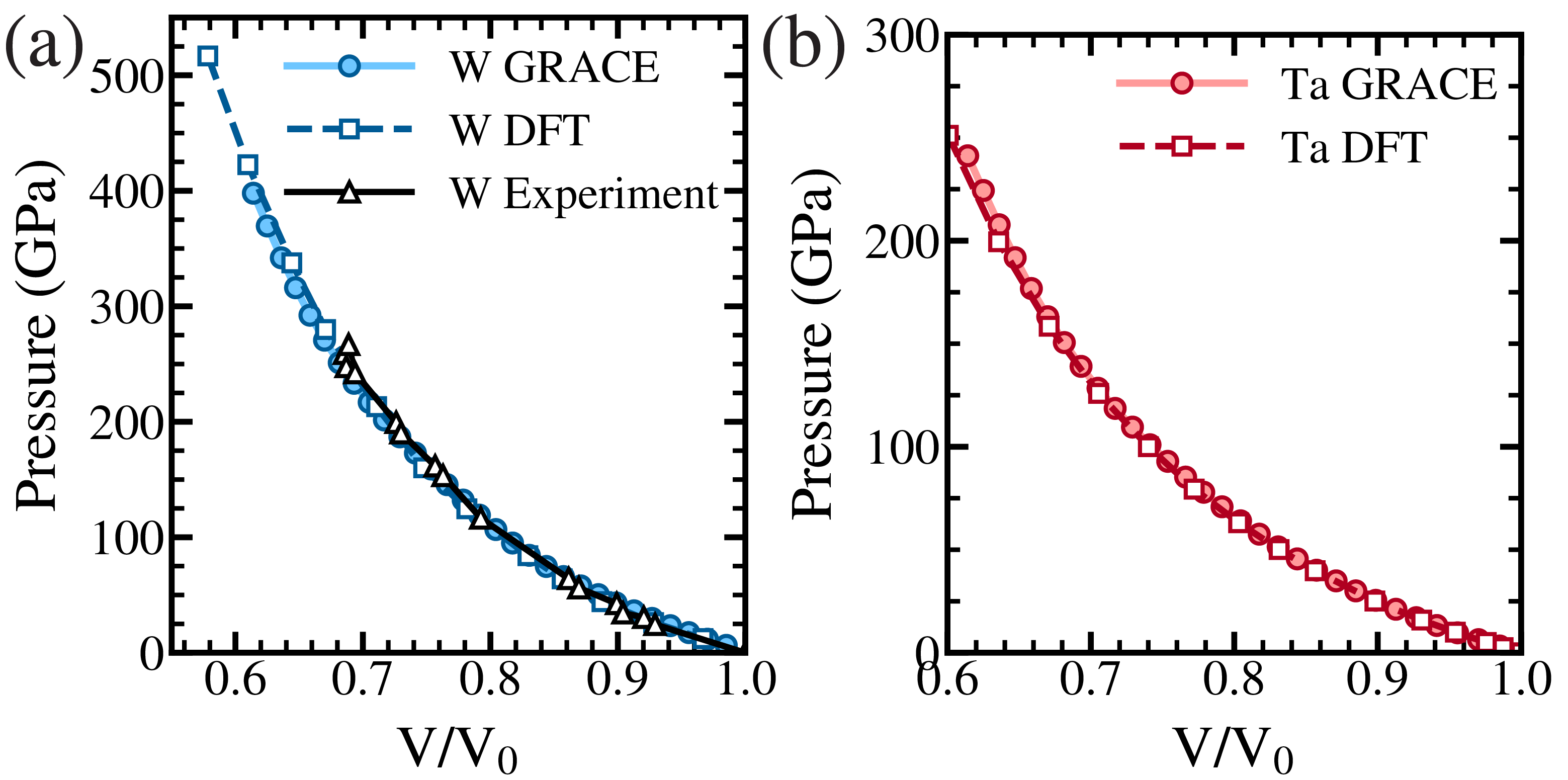}
\caption{High-pressure response of BCC (a) W and (b) Ta as a function of normalized volume, $V/V_0$. GRACE predictions are compared with DFT results for both elements and experimental data for W from Refs.~\cite{kinslow1970high,ravelo2013shock}.}
\label{fig:high-pressure}
\end{figure}


\subsubsection{Phonon Dispersion}
\begin{figure}[H]
\centering
\includegraphics[width=0.8\textwidth]{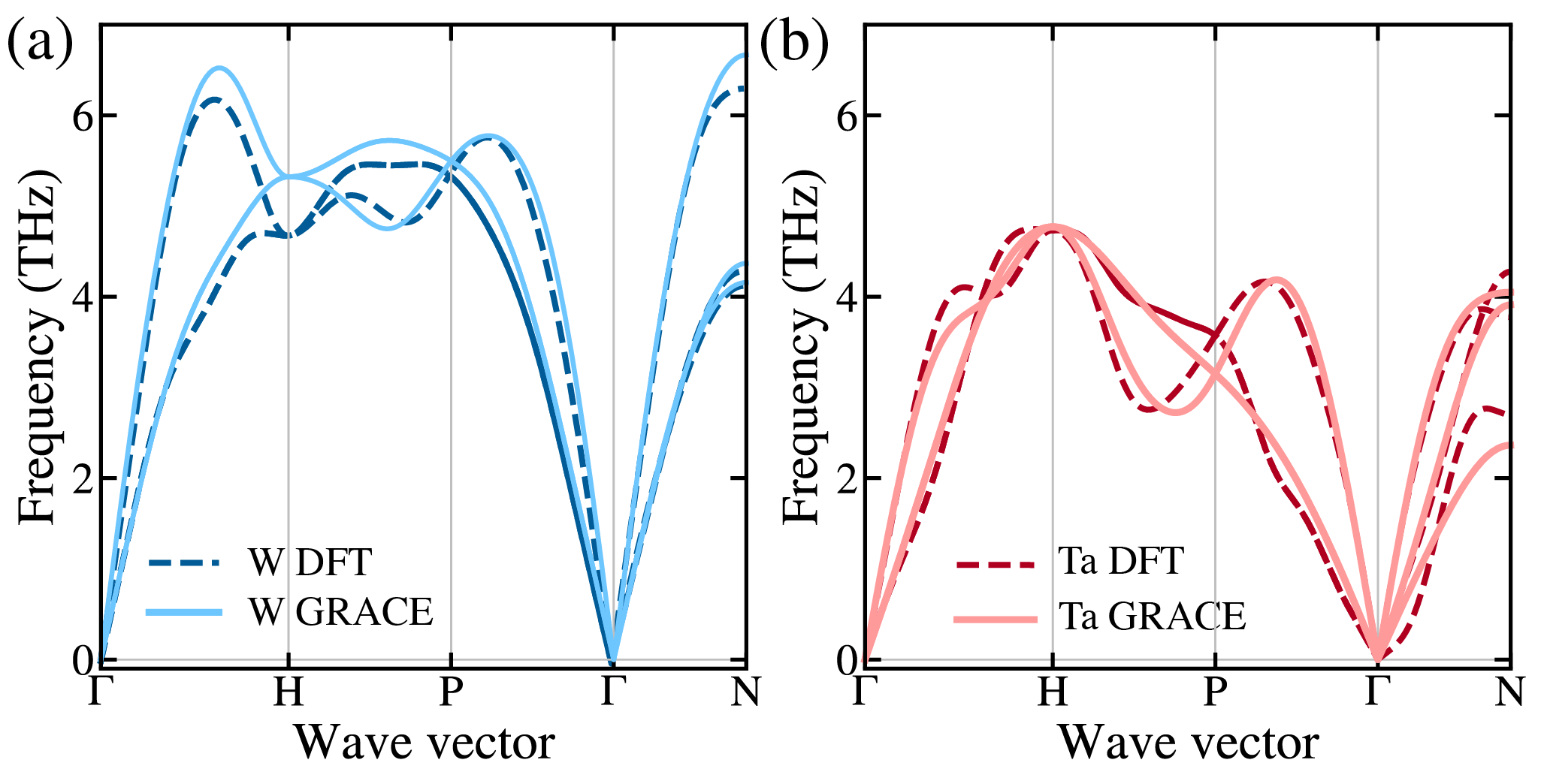}
\caption{Phonon dispersion relations of unary BCC (a) W and (b) Ta along the $\Gamma$--H--P--$\Gamma$--N high-symmetry path. The DFT curves were calculated in this study.}
\label{fig:phonon_dispersion}
\end{figure}

Phonon dispersion calculations were performed for unary BCC W and Ta using DFT and GRACE. Harmonic force constants were obtained using the finite-displacement method implemented in \texttt{Phonopy}~\cite{togo2015first}. For each method, the equilibrium lattice parameter was determined independently, and $4\times4\times4$ primitive-cell supercells were generated with atomic displacements of $0.01~\text{\AA}$. Static DFT calculations and zero-step LAMMPS calculations with GRACE were performed for the displaced structures, and the resulting forces were used to construct the force constants. Phonon frequencies were evaluated along the BCC high-symmetry path using $101$ wave-vector points per segment. As shown in Fig.~\ref{fig:phonon_dispersion}, GRACE closely reproduces the DFT dispersion of W and Ta.

\subsubsection{Generalized Stacking Fault Energy}
\begin{figure}[H]
    \centering
    \includegraphics[width=0.55\textwidth]{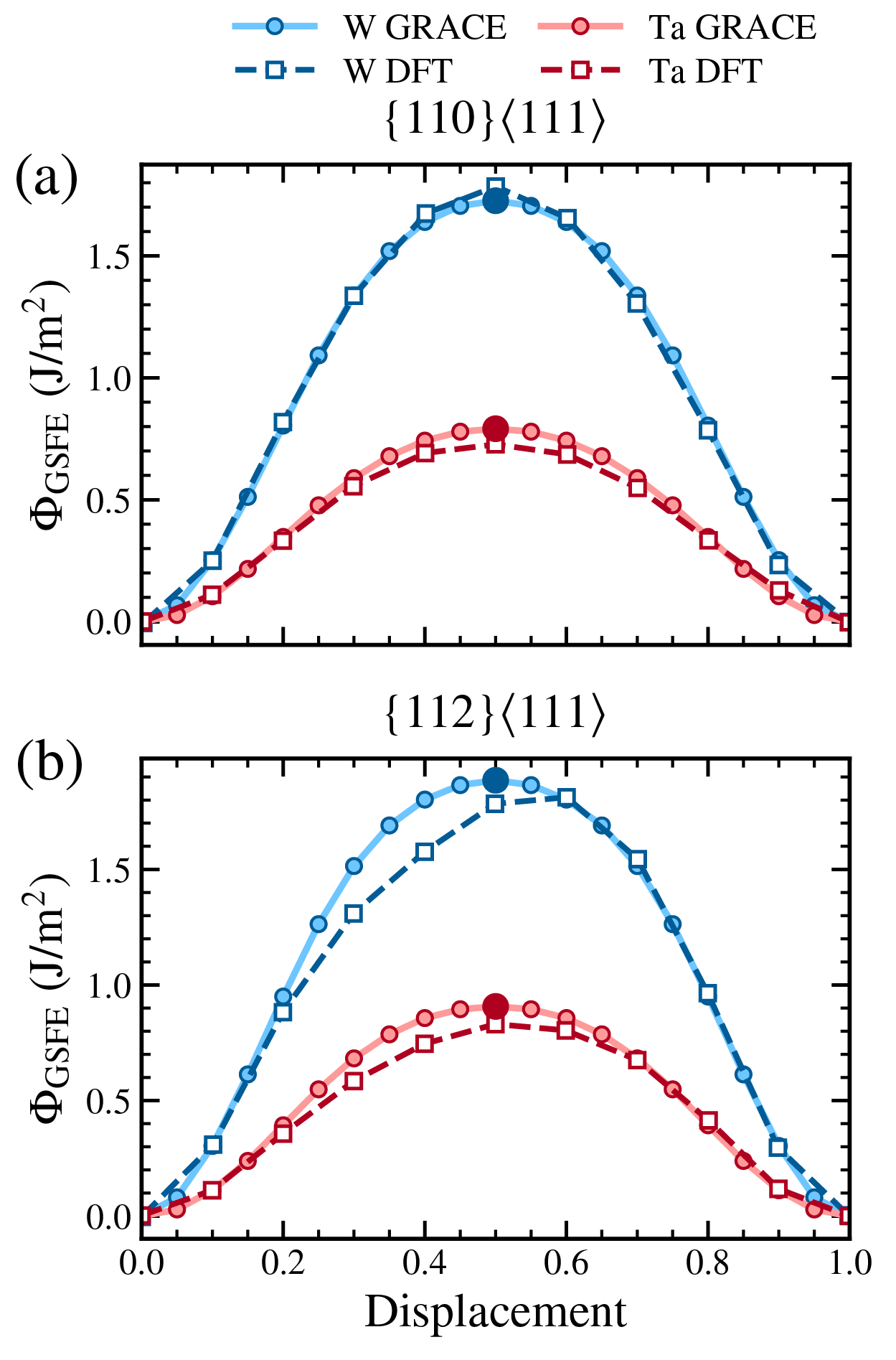}
    \caption{Generalized stacking fault energy curves for representative BCC slip systems in W and Ta along the (a) $\{110\}\langle111\rangle$ (b) $\{112\}\langle111\rangle$. The DFT data are taken from Ref.~\cite{zhang2023unstable}.}
    \label{fig:gsfe_curves}
\end{figure}

Generalized stacking fault energy (GSFE) curves were computed for the BCC $\{110\}\langle111\rangle$, $\{110\}\langle110\rangle$, $\{110\}\langle100\rangle$, $\{211\}\langle110\rangle$, and $\{211\}\langle111\rangle$ slip systems in W and Ta. For each case, a periodic $6\times6\times24$ oriented supercell was divided into two halves, with the upper half displaced over one lattice translation using 21 increments. Atomic relaxation was restricted to the fault-plane normal, and the excess energy was normalized by the area of the two periodic faults. The maximum GSFE values and corresponding normalized displacements are reported in Table~\ref{tab:gsfe}; all barriers occur near $u=0.50$. Figure~\ref{fig:gsfe_curves} compares the predicted GSFE curves with available DFT data for representative $\{110\}\langle 111 \rangle$ and $\{112\}\langle 111 \rangle$ slip systems.

\begin{table}[H]
\centering
\caption{Generalized stacking fault energy barriers for representative BCC slip systems in W and Ta.}
\label{tab:gsfe}
\footnotesize
\setlength{\tabcolsep}{3.5pt}
\renewcommand{\arraystretch}{0.92}
\resizebox{0.7\textwidth}{!}{%
\begin{tabular}{lcccc}
\toprule
\multirow{2}{*}{Slip plane/direction}
& \multicolumn{2}{c}{W, $\Phi_{\mathrm{GSFE}}^{\mathrm{max}}$ \si{\joule\per\meter\squared}}
& \multicolumn{2}{c}{Ta, $\Phi_{\mathrm{GSFE}}^{\mathrm{max}}$ \si{\joule\per\meter\squared}} \\
\cmidrule(lr){2-3}
\cmidrule(lr){4-5}
& DFT & GRACE & DFT & GRACE \\
\midrule
$\{110\}\langle 111 \rangle$
& 1.786\textsuperscript{a} & 1.726
& 0.704\textsuperscript{b}; 0.840\textsuperscript{c} & 0.791 \\

$\{110\}\langle 110 \rangle$
& -- & 3.388
& 1.628\textsuperscript{b} & 1.807 \\

$\{110\}\langle 100 \rangle$
& -- & 3.388
& 1.628\textsuperscript{b}; 1.951\textsuperscript{c} & 1.807 \\

$\{211\}\langle 110 \rangle$
& -- & 4.684
& 3.330\textsuperscript{b} & 3.566 \\

$\{211\}\langle 111 \rangle$
& 1.813\textsuperscript{a} & 1.886
& 0.844\textsuperscript{b}; 1.000\textsuperscript{c} & 0.907 \\
\bottomrule
\end{tabular}%
}

\vspace{2pt}
\begin{minipage}{0.82\textwidth}
\scriptsize
\textsuperscript{a} Ref.~\cite{zhang2023unstable};
\textsuperscript{b} Ref.~\cite{lin2022development};
\textsuperscript{c} Ref.~\cite{ravelo2013shock}.
\end{minipage}
\end{table}

\subsubsection{Transformation Paths}
\label{subsec:transformation-paths}
\begin{figure}[H]
    \centering
    \includegraphics[width=0.50\textwidth]{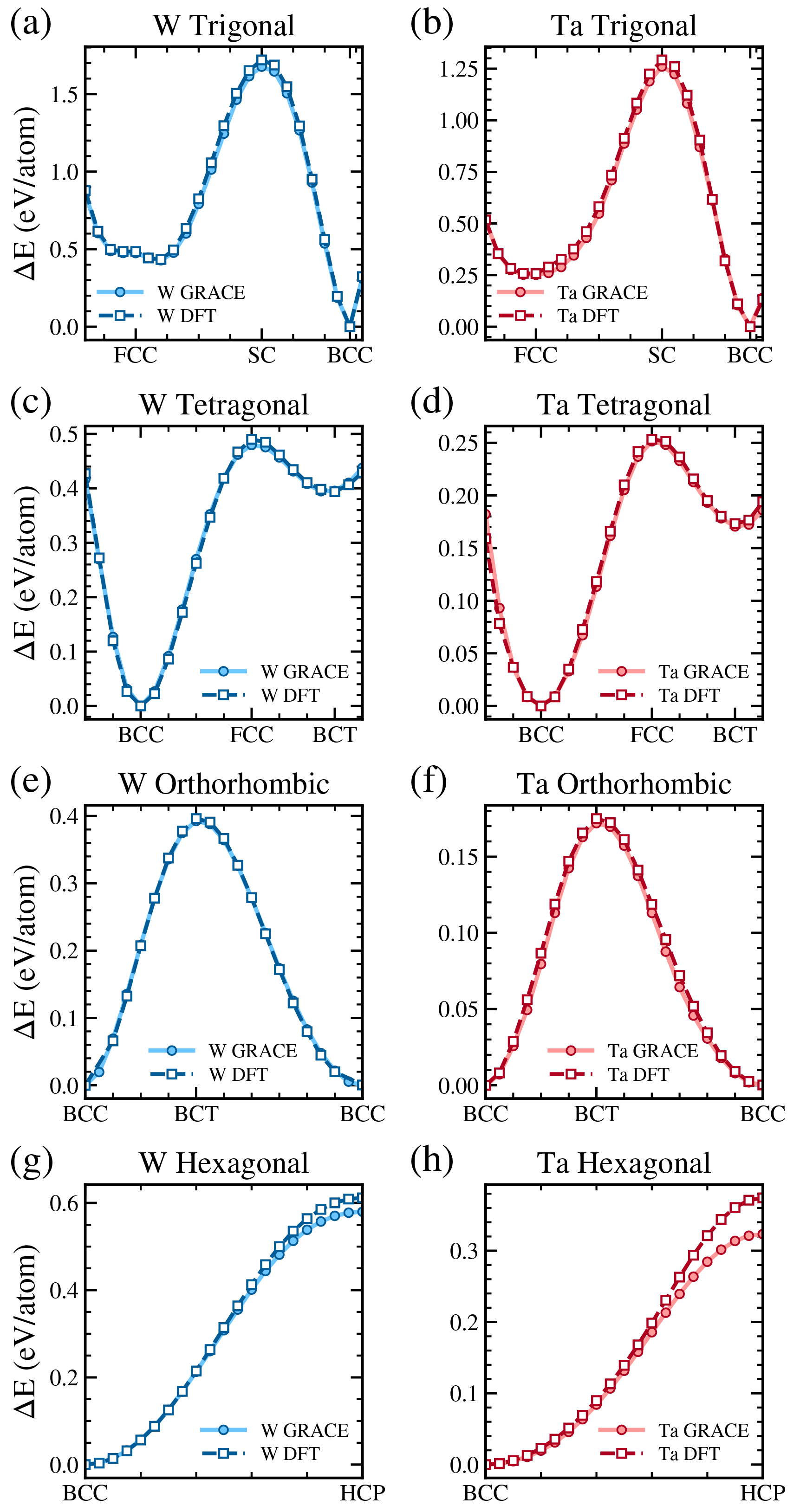}
    \caption{
    Relative energy per atom, $\Delta E$, along homogeneous transformation paths for W and Ta: (a,b) trigonal, (c,d) tetragonal, (e,f) orthorhombic, and (g,h) hexagonal deformation paths. GRACE predictions are compared with DFT calculations performed in this work.
    }
    \label{fig:transformation_paths}
\end{figure}

Transformation pathways were used to assess whether the GRACE potential reproduces the energetic response of W and Ta as their BCC structures are distorted along trigonal, tetragonal, hexagonal, and orthorhombic paths. Prescribed homogeneous distortions were applied to periodic $6\times6\times6$ cells, and the potential energy per atom was evaluated without atomic relaxation. The trigonal path varied the rhombohedral angle from $50^\circ$ to $112^\circ$, while the tetragonal Bain path varied $c/a$ from 0.8 to 1.8 at constant volume. The hexagonal Burgers-like path transformed BCC toward HCP through coupled lattice distortion and atomic-layer shuffle, parameterized by $\lambda=0$ to 1. The orthorhombic path varied $c/a$ from $1/\sqrt{2}$ to $\sqrt{2}$ at constant volume, connecting two symmetry-equivalent BCC variants through a body-centered tetragonal configuration. The resulting GRACE energies were compared with DFT reference calculations performed in this work. As shown in Fig.~\ref{fig:transformation_paths}, GRACE agrees well with DFT for both W and Ta across all transformation paths.

\subsubsection{Vacancy Migration Energy}
\label{subsubsec:vacancy_migration_barrier_cal}
\begin{figure}[H]
\centering
\includegraphics[width=0.4\textwidth]{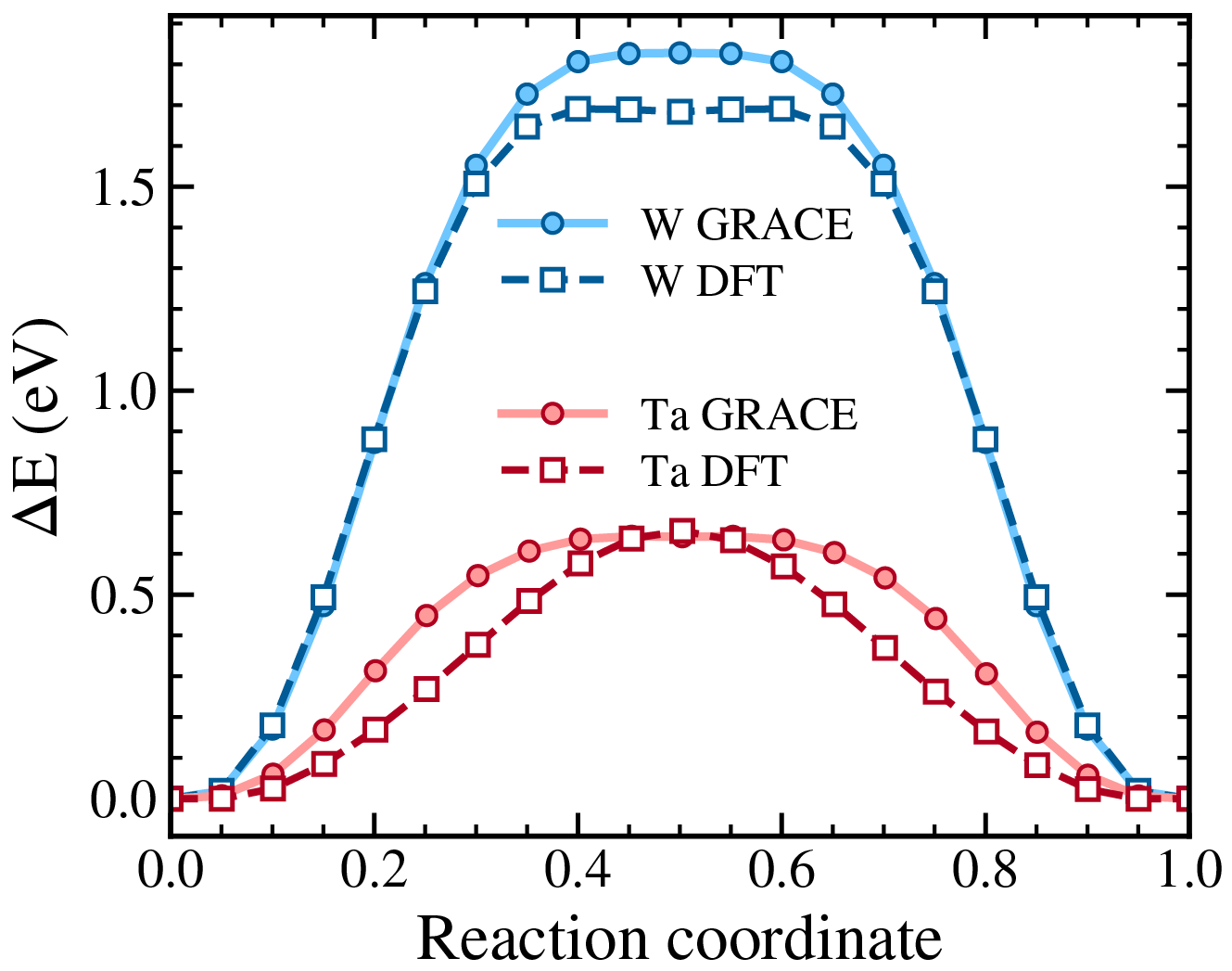}
\caption{Vacancy migration energy profiles for BCC W and Ta from GRACE and DFT. W has a higher migration barrier of about $1.7$--$1.8~\mathrm{eV}$, while Ta has a lower barrier of about $0.65~\mathrm{eV}$.}
\label{fig:vacancy_migration}
\end{figure}

Vacancy migration barriers were calculated using NEB for a nearest-neighbor $a_0/2\langle111\rangle$ vacancy hop in periodic $6\times6\times6$ BCC supercells. The relaxed endpoints were connected by 21 replicas and optimized using the quick-min algorithm with a spring constant of $1.0$ and a force tolerance of $10^{-8}~\mathrm{eV/\AA}$. The barrier was taken as the maximum energy relative to the initial state. As shown in Fig.~\ref{fig:vacancy_migration}, GRACE predicts barriers of approximately $1.8~\mathrm{eV}$ for W and $0.65~\mathrm{eV}$ for Ta, compared with DFT values of about $1.7~\mathrm{eV}$ and $0.65~\mathrm{eV}$, respectively.

\subsubsection{Di-Vacancy Binding Energy}
\label{subsec:divacancy-binding-energy}
\begin{figure}[H]
\centering
\includegraphics[width=0.40\textwidth]{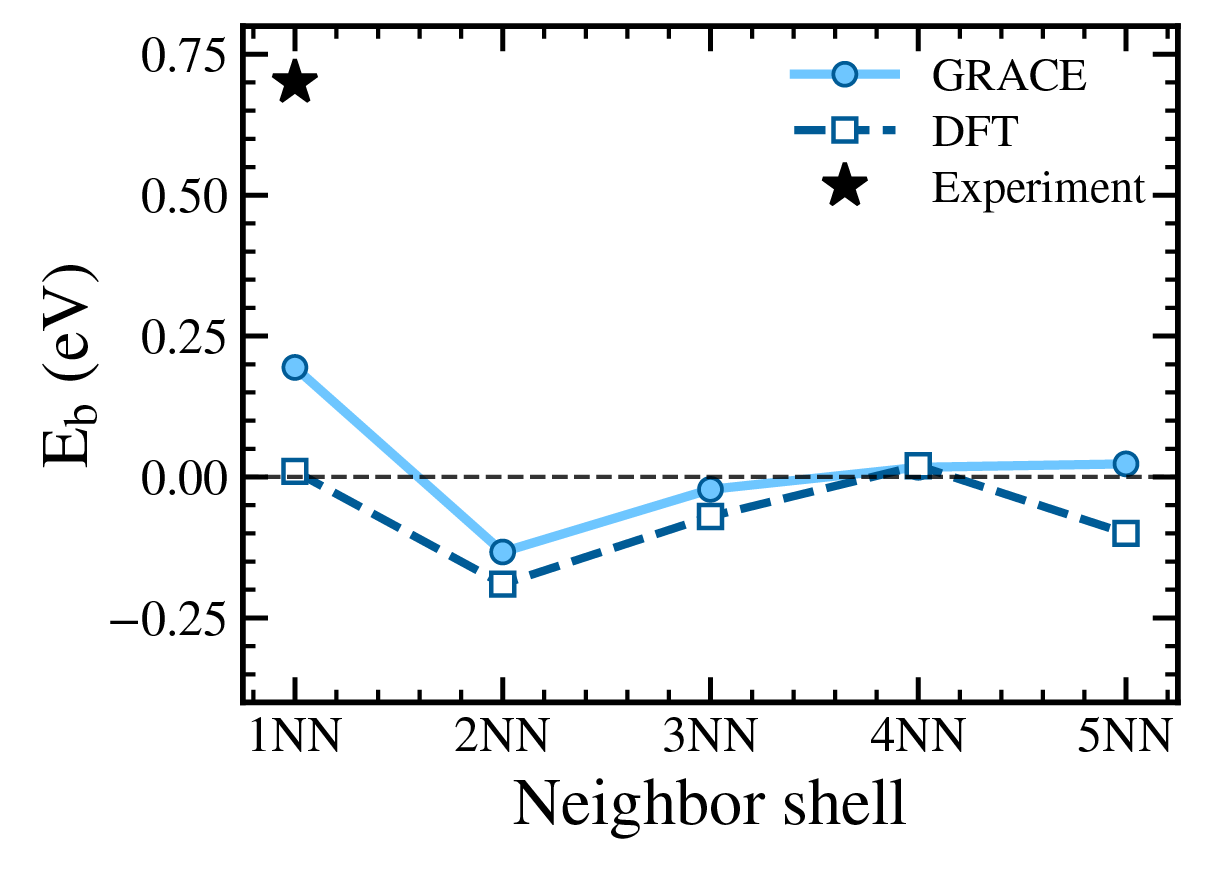}
\caption{Di-vacancy binding energy in BCC W as a function of vacancy-pair nearest-neighbor separation. GRACE predictions are compared with DFT data from Ref.~\cite{heinola2018stability} and the experimental 1NN value from Ref.~\cite{park1983quantitative}.}
\label{fig:di-vacancy}
\end{figure}

The di-vacancy binding energy was calculated for vacancy pairs in the first through fifth nearest-neighbor shells of BCC W. A periodic $5\times5\times5$ conventional BCC supercell was used, and the equilibrium lattice parameter was first obtained by relaxing the bulk cell. Single- and di-vacancy structures were then relaxed using conjugate-gradient minimization with anisotropic zero-pressure box relaxation. All symmetry-equivalent vacancy-pair configurations within each neighbor shell were evaluated, and the binding energy was calculated as $E_b=2E_{\mathrm{vac}}-E_{\mathrm{di-vac}}-E_{\mathrm{bulk}},$ where positive values indicate attractive binding. GRACE was compared with DFT data from Ref.~\cite{heinola2018stability} and the experimental 1NN value from Ref.~\cite{park1983quantitative}. The 4NN and 5NN DFT results should be interpreted cautiously because they were obtained using a relatively small $5\times5\times5$ supercell~\cite{heinola2018stability}. As shown in Fig.~\ref{fig:di-vacancy}, GRACE captures the main DFT trend, including the repulsive 2NN interaction, weak long-range binding, and an attractive 1NN interaction, although its magnitude is underestimated relative to experiment. We also note that different DFT studies report significantly varying values  ~\cite{heinola2018stability} , potentially indicating subtle electronic effects.

\subsubsection{Atomic String Displacement Curves}

The atomic string-displacement model was used to probe the local energy response associated with $1/2\langle111\rangle$ screw dislocations in BCC W and Ta, following Gilbert and Dudarev~\cite{gilbert2010ab,marinica2013interatomic}. Periodic $4\times4\times4$ cells were oriented with $z\parallel[111]$, and the equilibrium structure was first relaxed to zero pressure using conjugate-gradient minimization. Three displacement modes were considered: a single $[111]$ atomic string, two neighboring strings displaced together, and two neighboring strings displaced in opposite directions to produce shear. For each mode, the selected strings were rigidly displaced along $[111]$ over one Burgers-vector magnitude, $b=a_0\sqrt{3}/2$, using 41 equally spaced increments, while all remaining atoms were kept fixed and no relaxation was performed after displacement. The energy at each increment was evaluated relative to the relaxed reference configuration. 

\begin{figure}[H]
\centering
\includegraphics[width=0.30\textwidth]{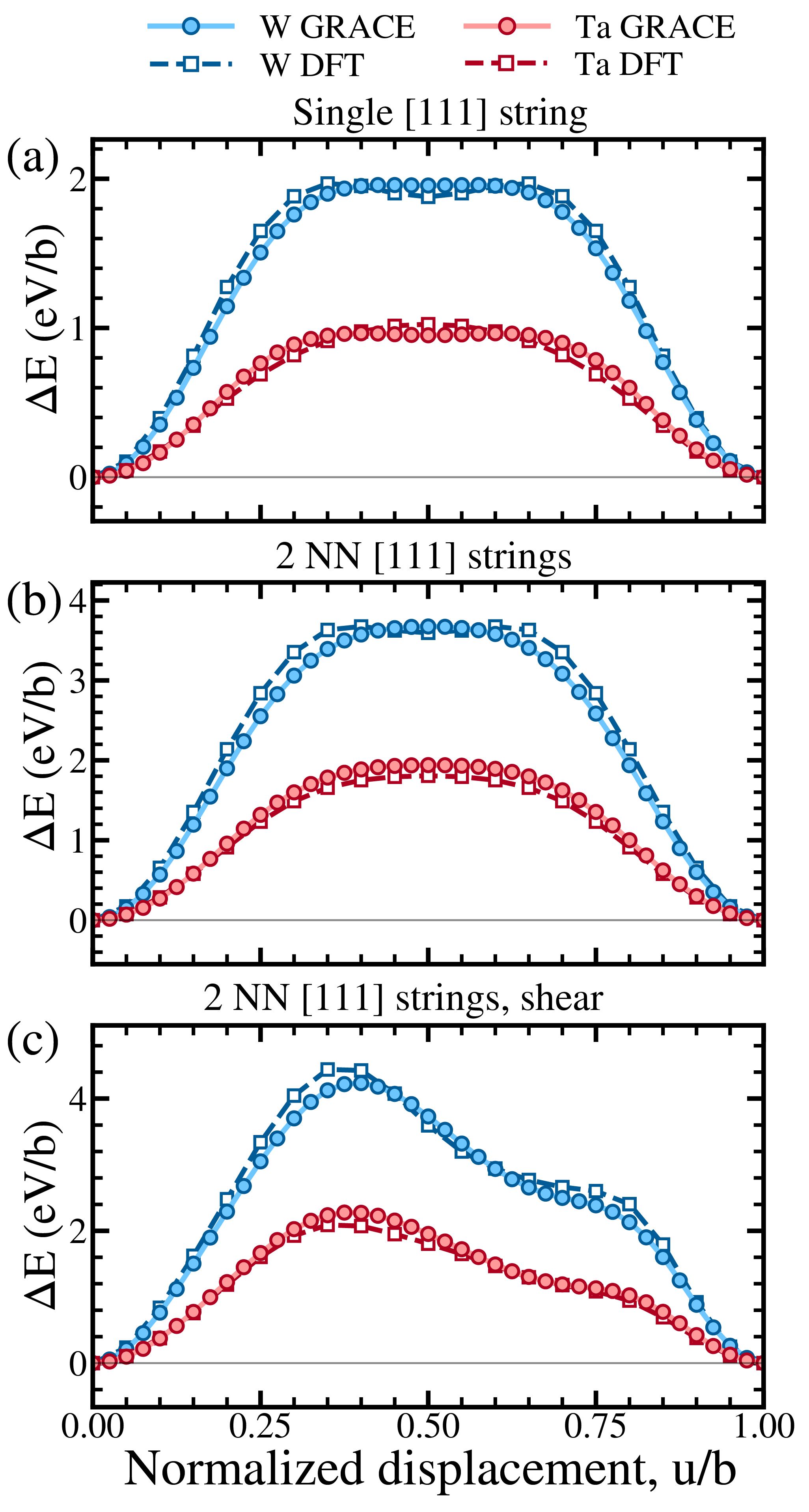}
\caption{Atomic string-displacement energies for BCC W and Ta from GRACE and DFT calculations from the present study. The panels show (a) single-string, (b) two-string collective, and (c) two-string shear displacements.}
\label{fig:string-displacement}
\end{figure}

As shown in Fig.~\ref{fig:string-displacement}, GRACE agrees extremely well with the DFT calculations for both W and Ta. The single-string barriers are approximately $2~\mathrm{eV}/b$ for W and $1~\mathrm{eV}/b$ for Ta, while the two-string collective barriers increase to about $3.5\text{--}3.7~\mathrm{eV}/b$ and $1.8\text{--}2.0~\mathrm{eV}/b$, respectively. The shear mode produces large barriers and asymmetric energy profiles, matching the DFT predictions.

\subsubsection{Differential Displacement Map}
\begin{figure}[H]
\centering
\includegraphics[width=0.6\textwidth]{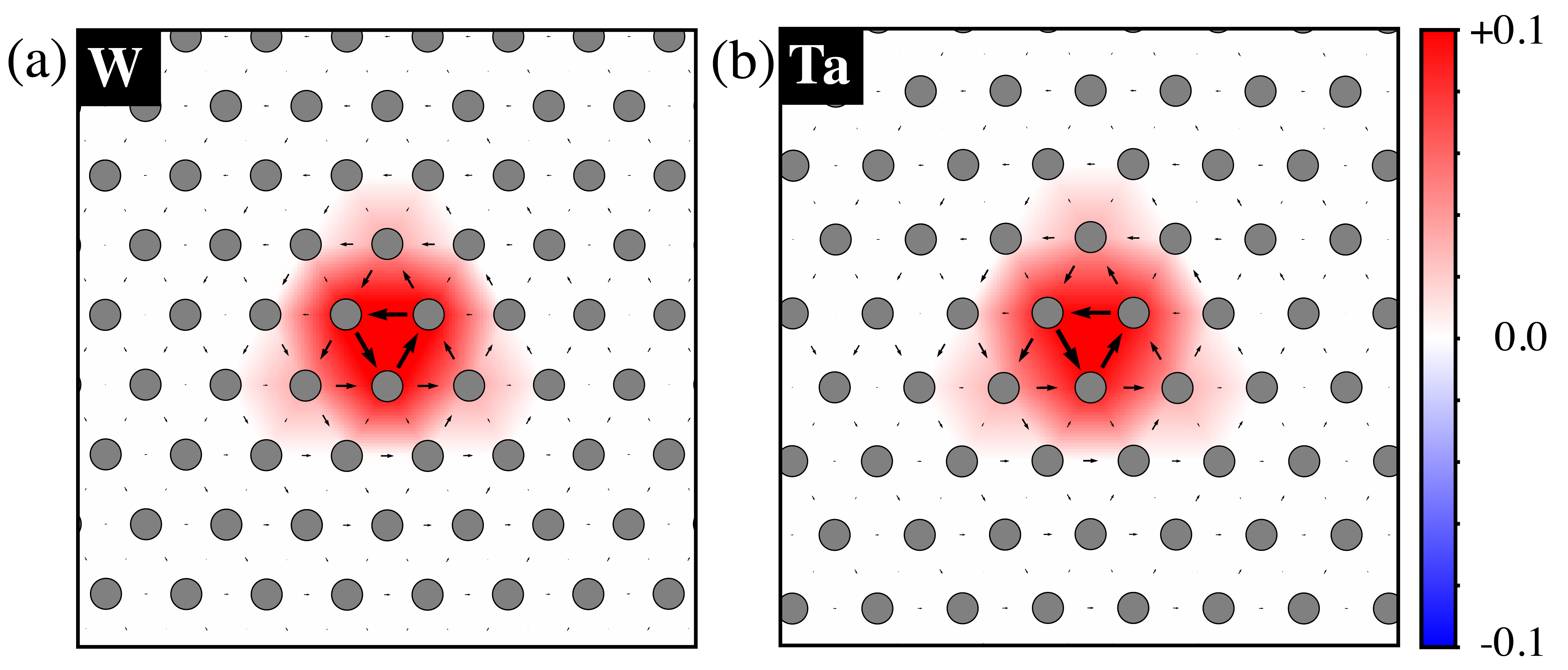}
\caption{Differential displacement maps of the $1/2\langle111\rangle$ screw-dislocation core in (a) W and (b) Ta predicted using GRACE. Both elements exhibit compact, localized, non-degenerate screw cores, consistent with the expected BCC transition-metal core structure.}
\label{fig:screw-ddm}
\end{figure}

Differential displacement maps were used to characterize the $1/2\langle111\rangle$ screw-dislocation cores in BCC W and Ta using the \texttt{atomman} library~\cite{hale2018interatomic}. Preconstructed reference and dislocated configurations were relaxed by conjugate-gradient minimization, with periodic boundaries along the dislocation line and shrink-wrapped boundaries in the transverse directions. Differential displacements were evaluated relative to the relaxed defect-free structure. As shown in Fig.~\ref{fig:screw-ddm}, GRACE predicts compact, localized, non-degenerate cores in both elements, with displacements concentrated near the core and rapidly decaying into the surrounding lattice.

\subsubsection{Peierls Energy}
\label{subsec:peierls-energy}
\begin{figure}[H]
    \centering
    \includegraphics[width=0.40\textwidth]{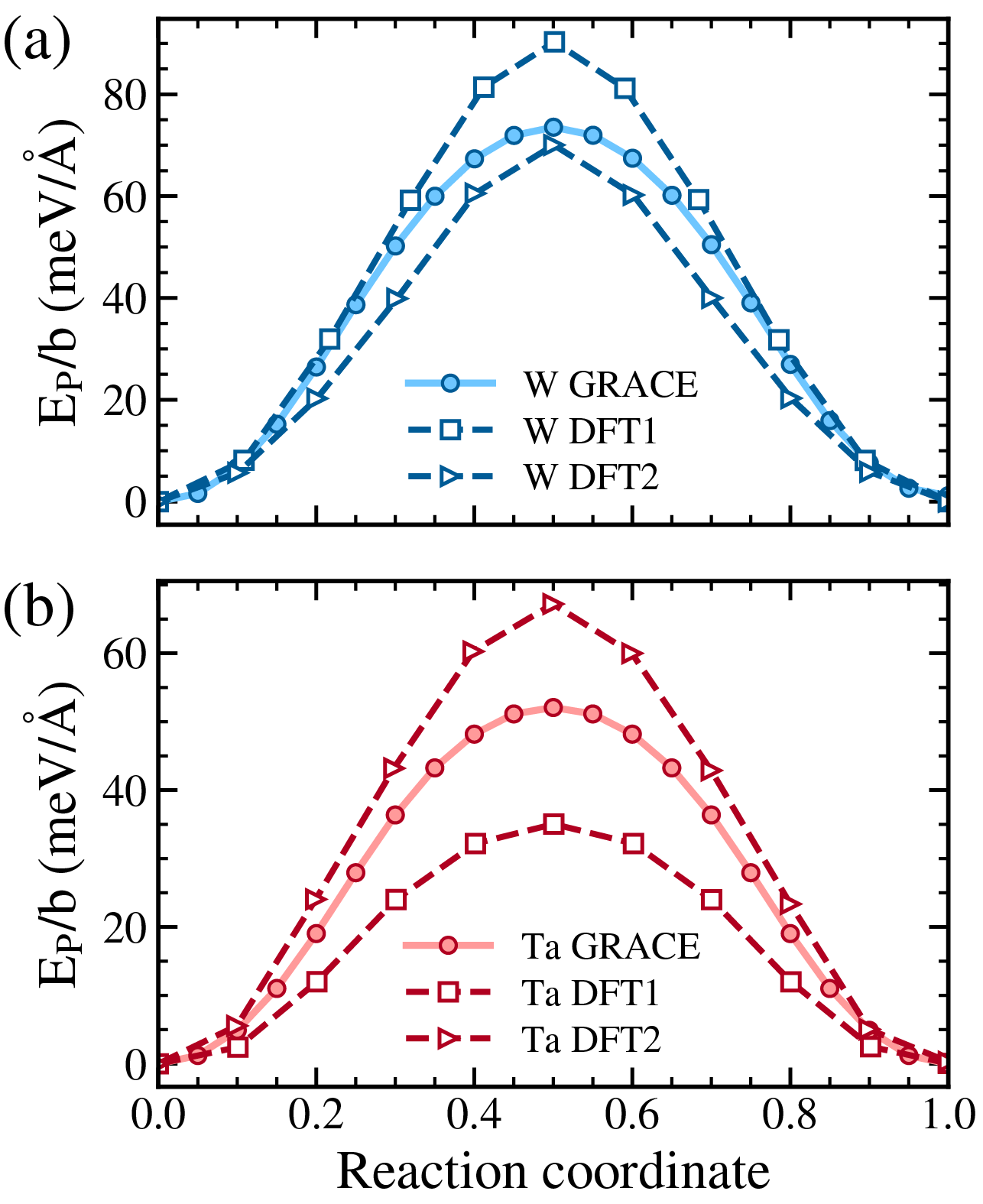}
    \caption{Normalized Peierls energy barrier, $E_{\mathrm{P}}/b$, as a function of reaction coordinate for $1/2\langle111\rangle$ screw-dislocation in (a) W and (b) Ta. GRACE predictions are compared with DFT data for W~\cite{marinica2013interatomic} and Ta~\cite{lin2022development,weinberger2013peierls}.}
    \label{fig:peierls-barrier}
\end{figure}

The Peierls barrier was calculated for $1/2\langle111\rangle$ screw-dislocation glide in BCC W and Ta using periodic dislocation dipoles generated with the \texttt{atomman} library~\cite{hale2018interatomic}. Two neighboring easy-core configurations were constructed on the $\{110\}$ slip plane, relaxed to zero pressure, and connected by a 21-replica NEB calculation. The replicas were optimized using the FIRE algorithm. As shown in Fig.~\ref{fig:peierls-barrier}, GRACE captures the overall barrier shape and symmetry for both elements. For W, the predicted barrier follows the lower DFT reference, while for Ta it lies within the range of reported DFT values. Since the double-kink formation energy scales linearly with the square root of the Peierls barrier~\cite{swinburne2013theory,marinica2013interatomic}, this agreement is important for describing thermally activated screw-dislocation motion.

\subsection{Binary Validation}

\subsubsection{Lattice Constant}
The composition-dependent lattice constant was evaluated for random BCC W--Ta solid solutions using periodic $6\times6\times6$ supercells. Fifteen compositions spanning the full range from pure W to pure Ta were generated by randomly assigning Ta atoms to BCC lattice sites using a fixed seed. Each structure was initialized using the Vegard-law lattice constant and relaxed by conjugate-gradient minimization with anisotropic zero-pressure box relaxation. The equilibrium lattice constant was obtained from the relaxed cell length, while the pure W and Ta endpoints were calculated independently using the same procedure.

\begin{figure}[H]
\centering
\includegraphics[width=0.40\textwidth]{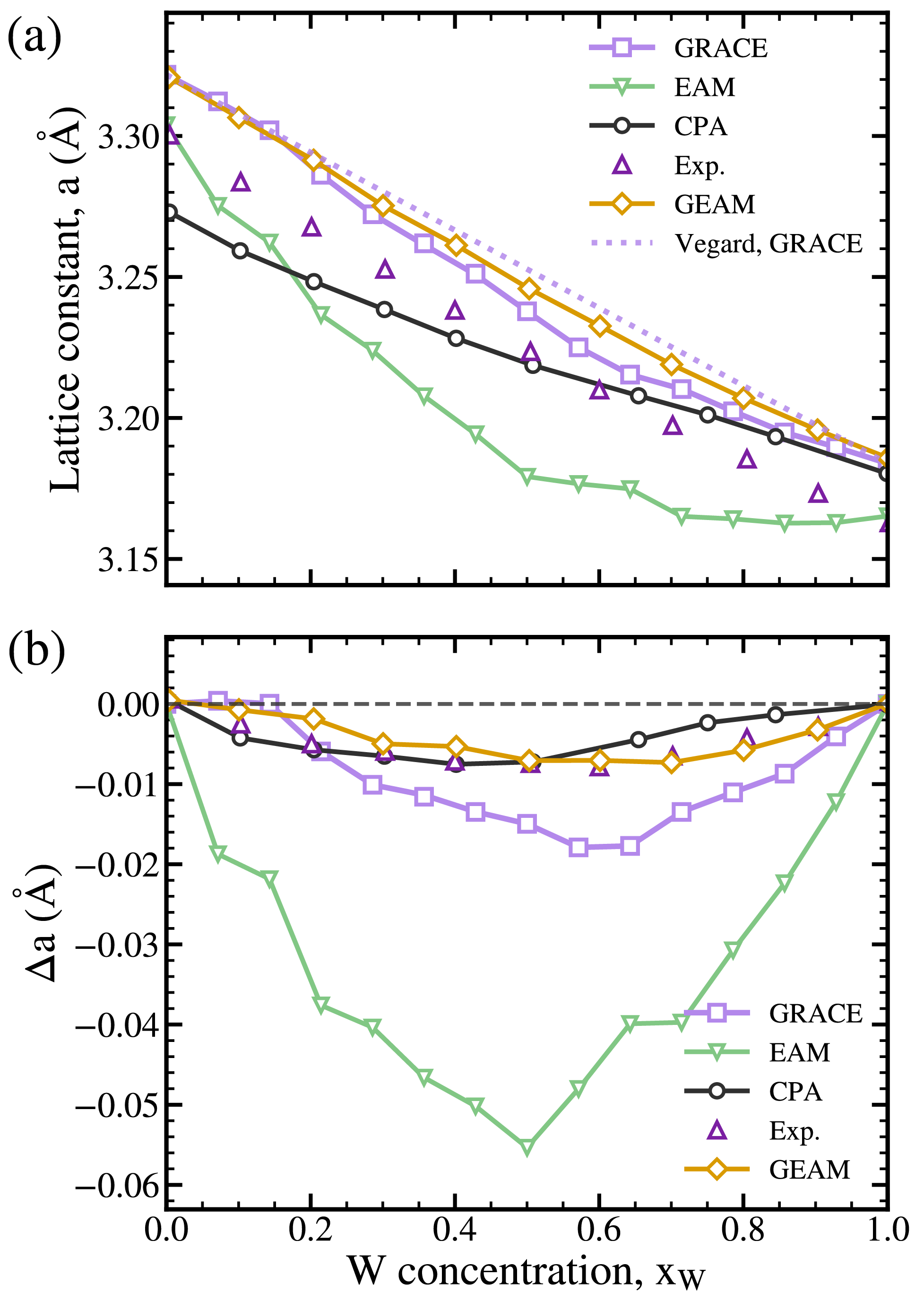}
\caption{Composition-dependent lattice constant of BCC W--Ta alloys. (a) Lattice constant, $a$, as a function of W concentration, $x_{\mathrm{W}}$, compared with Vegard's law. (b) Departure from Vegard's law, $\Delta a$. GRACE captures the monotonic decrease in lattice constant with increasing W content and predicts a modest negative deviation from Vegard's law, in closer agreement with CPA~\cite{turchi2001first} and experiment~\cite{krishnan1987tantalum} than EAM~\cite{chen2019development}.}
\label{fig:lattice-constant}
\end{figure}

Figure~\ref{fig:lattice-constant} compares the lattice constant as a function of W concentration, $x_{\mathrm{W}},$ predicted by GRACE, EAM~\cite{chen2019development}, GEAM~\cite{sharma2023development}, CPA~\cite{turchi2001first}, and experiment~\cite{krishnan1987tantalum}. The lattice constant decreases monotonically with increasing W content, consistent with the smaller equilibrium lattice constant of BCC W. GRACE remains close to the CPA and experimental data across the full composition range, whereas EAM underestimates the lattice constant at intermediate compositions.

The deviation from Vegard's law, $\Delta a,$ shown in Fig.~\ref{fig:lattice-constant}(b), indicates a very weak negative bowing. GRACE reproduces this slight contraction relative to ideal linear mixing, in qualitative agreement with CPA and experiment, although the deviation is somewhat larger near intermediate compositions.

\subsubsection{Elastic Constants and Bulk Modulus}
Elastic constants and the bulk modulus were evaluated for random BCC W--Ta alloys using periodic $6\times6\times6$ supercells at 15 compositions. Each structure was initialized from Vegard’s law, randomly populated, and relaxed to zero pressure. The elastic tensor was obtained from central finite differences of the stress response to axial and shear strains of $\pm5\times10^{-4}$ without post-strain atomic relaxation. Cubic projection yielded $C_{11}$, $C_{12}$, and $C_{44}$, with $B=(C_{11}+2C_{12})/3$.

\begin{figure}[H]
\centering
\includegraphics[width=0.30\textwidth]{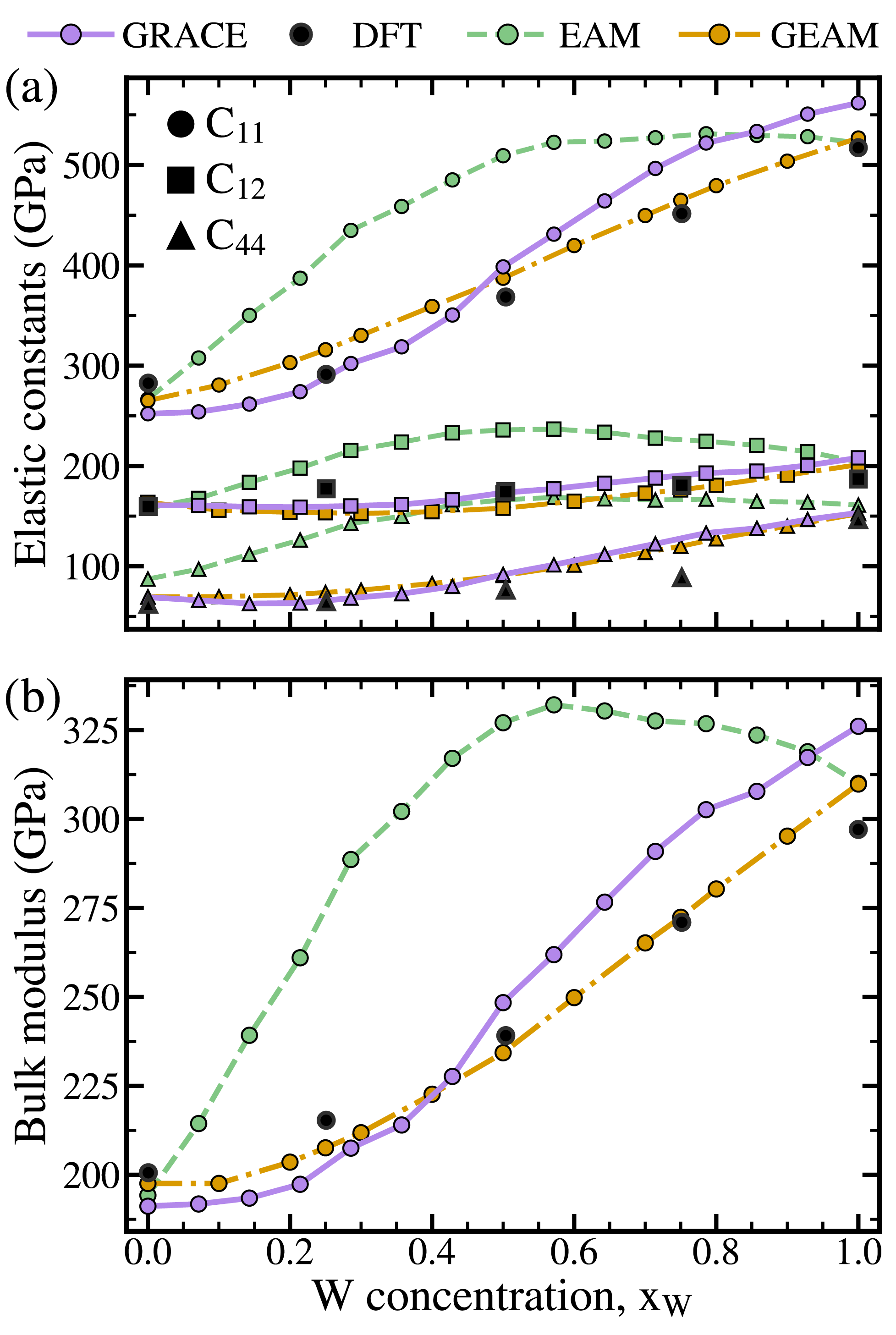}
\caption{Elastic constants and bulk modulus of BCC W--Ta alloys as a function of W concentration, $x_{\mathrm{W}}$. (a) Elastic constants $C_{11}$, $C_{12}$, and $C_{44}$ predicted by GRACE, EAM~\cite{chen2019development}, and GEAM~\cite{sharma2023development} compared with DFT~\cite{wei2014first} data. (b) Bulk modulus predicted by the same models and compared with DFT~\cite{wei2014first}. }
\label{fig:elastic-constants-bulk-modulus}
\end{figure}

Figure~\ref{fig:elastic-constants-bulk-modulus} compares GRACE, EAM~\cite{chen2019development}, and GEAM~\cite{sharma2023development} predictions with DFT~\cite{wei2014first} data as a function of W concentration, $x_{\mathrm{W}}$. GRACE captures the main DFT trends: $C_{11}$ and $C_{44}$ increase with W content, while $C_{12}$ varies more weakly. The bulk modulus also increases monotonically with W concentration, and GRACE follows the DFT trend more closely than EAM, which overestimates it over much of the composition range.

\subsubsection{Point Defect Energetics in W--Ta}

The energetics of Ta-related point defects in bcc W were evaluated to assess
the ability of GRACE to describe solute--defect interactions relevant to
radiation damage and microstructural evolution. The calculated formation and
binding energies are summarized in Table~\ref{tab:Ta_defect_energetics}.

\begin{table}[H]
\centering
\caption{Energetics of Ta point defects in bcc W from DFT calculations compared with GRACE, EAM~\cite{chen2019development}, and GEAM~\cite{sharma2023development}. All energies are in eV.}
\label{tab:Ta_defect_energetics}
\begin{threeparttable}
\scriptsize
\setlength{\tabcolsep}{3pt}
\renewcommand{\arraystretch}{1.2}
\resizebox{0.6\textwidth}{!}{
\begin{tabular}{lcccc}
\hline
Property & DFT & GRACE & EAM & GEAM \\
\hline
$E^{f}_{\mathrm{sub}}(\mathrm{Ta})$ & $-0.45^{a,b}$, $-0.47^{c,d,e}$ & $-0.44$ & $-0.47$ & $-0.48$ \\
$E^{b}_{1\mathrm{nn}}(\mathrm{Ta{-}Ta})$ & $-0.09^{a}$, $-0.08^{b}$ & $-0.08$ & $-0.13$ & $-0.12$ \\
$E^{b}_{2\mathrm{nn}}(\mathrm{Ta{-}Ta})$ & $-0.04^{a}$, $-0.02^{b}$ & $-0.03$ & $-0.02$ & $-0.05$ \\
$E^{b}_{1\mathrm{nn}}(\mathrm{Ta{-}Vac})$ & $-0.04^{a,f}$, $-0.05^{b}$ & $-0.01$ & $-0.06$ & $0.03$ \\
$E^{b}_{2\mathrm{nn}}(\mathrm{Ta{-}Vac})$ & $-0.21^{a,b}$, $-0.20^{f}$ & $-0.12$ & $-0.03$ & $-0.10$ \\
$E^{b}_{\langle111\rangle}(\mathrm{Ta{-}SIA})$ & $-0.64^{c,e}$, $-0.57^{b}$ & $-0.44$ & $-0.59$ & $-0.05$ \\
$E^{f}_{\langle111\rangle}(\mathrm{W{-}Ta})$ & $10.38^{c}$, $10.01^{e}$ & $10.39$ & $9.69$ & $9.15$ \\
$E^{f}_{\langle110\rangle}(\mathrm{W{-}Ta})$ & $10.93^{c}$, $10.45^{e}$ & $11.14$ & $9.75$ & $10.18$ \\
$E^{f}_{\langle100\rangle}(\mathrm{W{-}Ta})$ & $11.36^{c}$ & $11.53$ & $10.56$ & $10.78$ \\
\hline
\end{tabular}
}
\begin{tablenotes}\scriptsize \item[] $^{a}$\cite{you2017clustering}; $^{b}$\cite{kong2014first}; $^{c}$\cite{setyawan2017ab}; $^{d}$\cite{becquart2012solute}; $^{e}$\cite{gharaee2015first}; $^{f}$\cite{becquart2012solute}. \end{tablenotes}

\end{threeparttable}
\end{table}

The formation energy of a substitutional Ta atom in BCC W was calculated as $E^{f}_{\mathrm{sub}}(\mathrm{Ta})=E_{\mathrm{Ta;sub}}-E_{\mathrm{W}}+\mu_{\mathrm{W}}-\mu_{\mathrm{Ta}},$ where $E_{\mathrm{W}}$ is the energy of the perfect W supercell and $\mu_{\mathrm{W}}$ and $\mu_{\mathrm{Ta}}$ are the elemental chemical potentials. GRACE predicts $E^{f}_{\mathrm{sub}}=-0.44~\mathrm{eV},$ in close agreement with the reported DFT range of approximately $-0.45$ to $-0.47~\mathrm{eV}.$ The negative value indicates that substitutional Ta is thermodynamically favored in the W matrix relative to the selected elemental reference states.

With the binding-energy convention used here, positive values indicate attractive interactions, whereas negative values indicate repulsive interactions. Ta--Ta interactions were quantified using $E^{b}_{\mathrm{Ta-Ta}}=2E^{f}_{\mathrm{sub}}-E^{f}_{\mathrm{2Ta}},$ where $E^{f}_{\mathrm{2Ta}}$ is the formation energy of a supercell containing two substitutional Ta atoms. GRACE predicts binding energies of $-0.08$ and $-0.03~\mathrm{eV}$ for first- and second-nearest-neighbor pairs, respectively, consistent with the corresponding DFT ranges of approximately $-0.08$ to $-0.09~\mathrm{eV}$ and $-0.02$ to $-0.04~\mathrm{eV}.$ GRACE therefore reproduces both the weak interaction strength and its decay with separation.

The Ta--vacancy binding energy was calculated as $E^{b}_{\mathrm{Ta-Vac}}=E_{\mathrm{Ta;sub}}+E_{\mathrm{Vac}}-E_{\mathrm{Ta-Vac}}-E_{\mathrm{W}}.$ GRACE predicts values of $-0.01$ and $-0.12~\mathrm{eV}$ for the 1NN and 2NN configurations, respectively. The 1NN result is close to the DFT range of approximately $-0.04$ to $-0.05~\mathrm{eV},$ whereas the 2NN interaction is weaker than the reported DFT range of approximately $-0.20$ to $-0.21~\mathrm{eV}$. GRACE correctly predicts stronger repulsion at 2NN than at 1NN, but underestimates its values at both separations.

The Ta--self-interstitial interaction was evaluated using $E^{b}_{\mathrm{Ta-SIA}}=E_{\mathrm{Ta;sub}}+E_{\mathrm{SIA}}-E_{\mathrm{Ta-SIA}}-E_{\mathrm{W}}.$ For the $\langle111\rangle$ configuration, GRACE predicts a binding energy of $-0.44~\mathrm{eV},$ compared with DFT values of approximately $-0.57$ to $-0.64~\mathrm{eV}$. Thus, GRACE reproduces the repulsive character of the Ta--SIA interaction but underestimates its magnitude.

Mixed W--Ta dumbbell formation energies were calculated as $E^{f}_{\mathrm{W-Ta}}=E_{\mathrm{defect}}-n_{\mathrm{W}}\mu_{\mathrm{W}}-n_{\mathrm{Ta}}\mu_{\mathrm{Ta}},$ where $E_{\mathrm{defect}}$ is the relaxed defect energy. GRACE predicts formation energies of $10.39,$ $11.14,$ and $11.53~\mathrm{eV}$ for the $\langle111\rangle,$ $\langle110\rangle,$ and $\langle100\rangle$ dumbbells, respectively. These values reproduce the DFT stability sequence $\langle111\rangle<\langle110\rangle<\langle100\rangle,$ identifying the $\langle111\rangle$ dumbbell as the most stable mixed-interstitial configuration.

\subsubsection{Ta Cluster Binding Energy}

The binding energy of Ta clusters was calculated to examine whether GRACE captures the tendency of Ta atoms to cluster in BCC W. For each cluster size, several compact Ta arrangements were relaxed, and the lowest-energy structure was used.

For clusters containing only Ta atoms, the binding energy was calculated as $E\sb{b}(\mathrm{Ta}\sb{n})=nE\sb{\mathrm{sol}}-E(\mathrm{Ta}\sb{n})-(n-1)E\sb{\mathrm{W}}$, where $E\sb{\mathrm{sol}}$ is the energy of a W supercell with one substitutional Ta atom, $E(\mathrm{Ta}\sb{n})$ is the energy of the supercell containing a Ta cluster of size $n$, and $E\sb{\mathrm{W}}$ is the energy of the pure W supercell. For vacancy-assisted clusters, the binding energy was calculated as $E\sb{b}(\mathrm{VacTa}\sb{n})=E\sb{\mathrm{vac}}+nE\sb{\mathrm{sol}}-E(\mathrm{VacTa}\sb{n})-nE\sb{\mathrm{W}}$, where $E\sb{\mathrm{vac}}$ is the energy of a W supercell containing one vacancy.

\begin{figure}[H]
\centering
\includegraphics[width=0.30\textwidth]{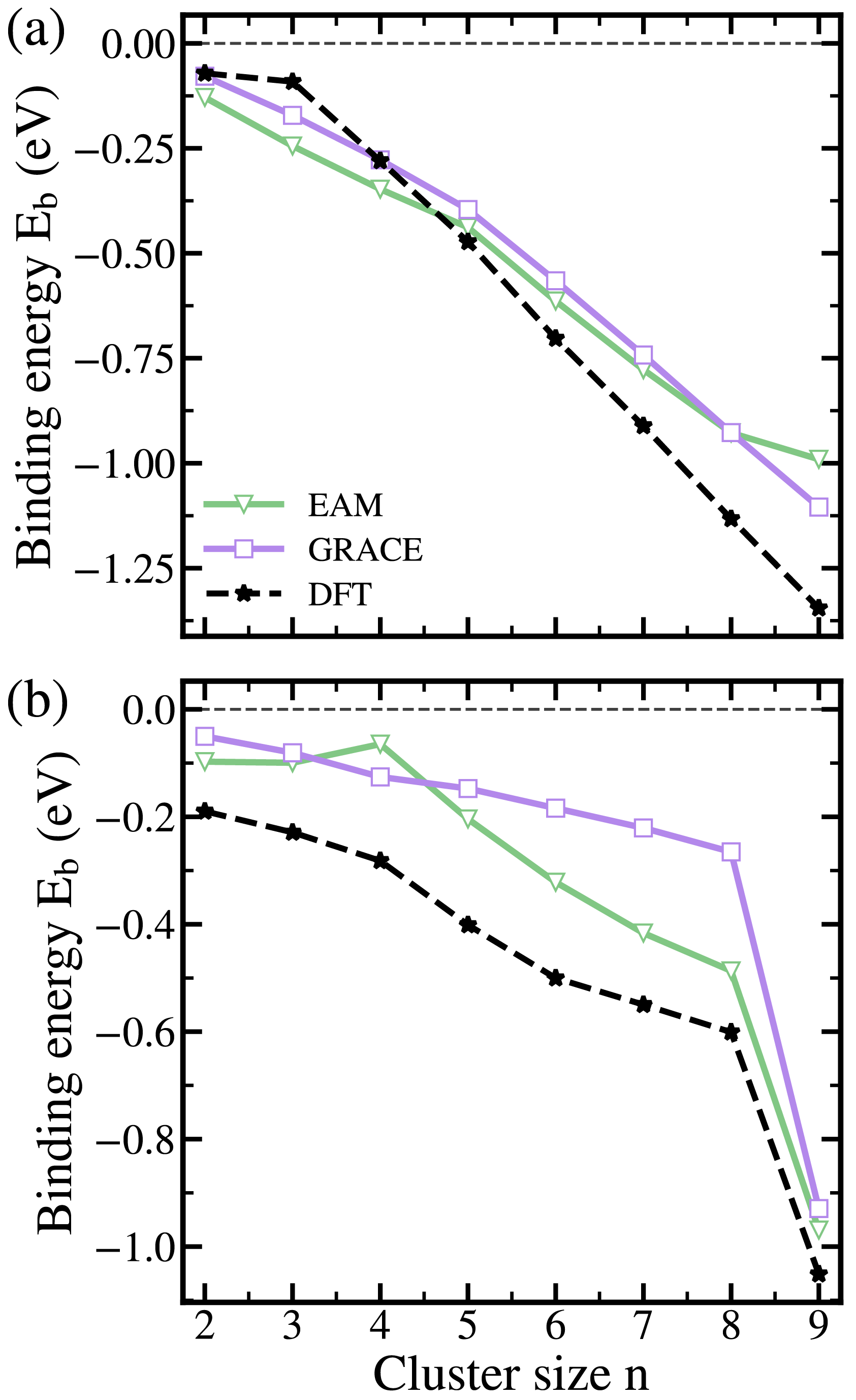}
\caption{Binding energies of (a) pure Ta clusters and (b) vacancy-assisted Ta clusters in BCC W for GRACE, EAM~\cite{chen2019development}, and DFT~\cite{you2017clustering}.}
\label{fig:ta-cluster-binding}
\end{figure}

Figure~\ref{fig:ta-cluster-binding} compares GRACE with EAM~\cite{chen2019development} and available DFT data~\cite{you2017clustering}. For both pure Ta clusters and vacancy-assisted clusters, the binding energies are negative and become more negative with increasing cluster size. With this sign convention, larger Ta clusters are less favorable. GRACE captures the overall DFT trend, although it predicts less negative binding energies, especially for larger clusters. Thus, GRACE slightly underestimates the strength of Ta--Ta and vacancy--Ta interactions, but correctly shows that Ta clustering in W is not favored. This agrees with previous calculations and irradiation experiments, which reported no clear Ta precipitation or clustering in W--Ta alloys~\cite{xu2017ion}.

\subsubsection{Enthalpy of Mixing of the W--Ta Binary Alloy}

The enthalpy of mixing of the W--Ta binary alloy was evaluated as a function of composition to assess whether the GRACE potential captures the thermodynamic tendency for alloy formation. For each composition, random substitutional BCC W--Ta alloy configurations were generated in a $10\times10\times10$ BCC supercell containing 2000 atoms. Ten independent configurations were generated using different random seeds, and each configuration was minimized with anisotropic box relaxation.
\begin{figure}[htbp]
\centering
\includegraphics[width=0.4\textwidth]{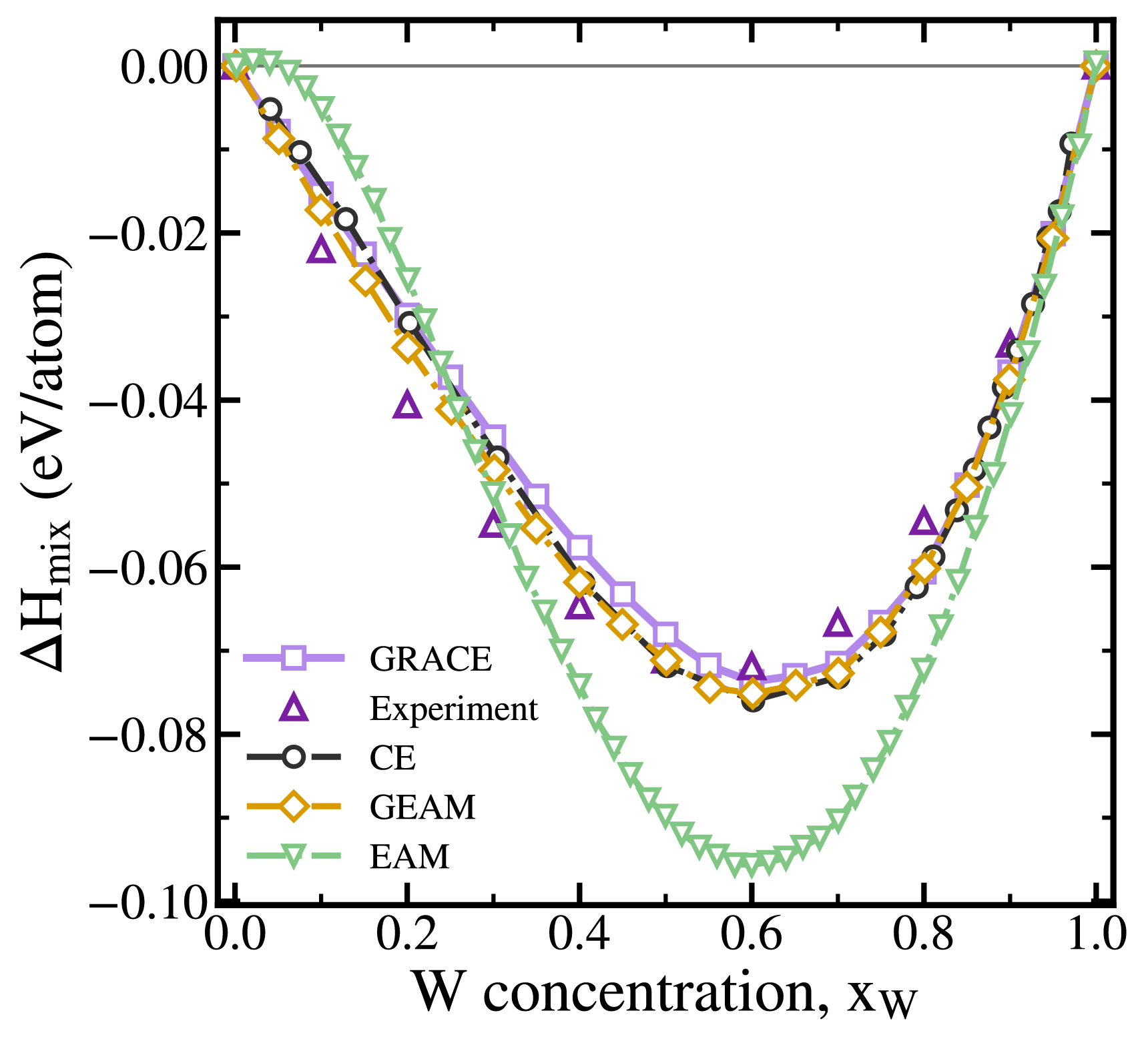}
\caption{
Enthalpy of mixing, $\Delta H_{\mathrm{mix}}$, of the W--Ta binary alloy as a function of W concentration, $x_{\mathrm{W}}$. GRACE predictions are compared with experimental data~\cite{blum2005prediction}, cluster expansion (CE)~\cite{muzyk2011phase}, GEAM~\cite{sharma2023development}, and EAM~\cite{chen2019development} results.
}
\label{fig:wta_hmix}
\end{figure}
The mixing enthalpy was computed as $ \Delta H_{\mathrm{mix}} = E_{\mathrm{WTa}} - \left[x_{\mathrm{Ta}}E_{\mathrm{Ta}}+(1-x_{\mathrm{Ta}})E_{\mathrm{W}}\right],$ where $E_{\mathrm{WTa}}$ is the relaxed alloy energy per atom, and $E_{\mathrm{W}}$ and $E_{\mathrm{Ta}}$ are the relaxed energies per atom of pure BCC W and BCC Ta calculated using the same potential. For each composition, the reported value corresponds to the lowest relaxed $\Delta H_{\mathrm{mix}}$ among the ten random configurations.

As shown in Fig.~\ref{fig:wta_hmix}, the reference data predict negative mixing enthalpies across most of the composition range, indicating that W--Ta alloy formation is energetically favorable relative to the pure BCC elements. The GRACE potential reproduces the composition dependence, including the negative values of the mixing enthalpy, $\Delta H_{\mathrm{mix}}$, the approach to zero near the pure W and Ta limits, and the minimum near intermediate W concentrations. GRACE agrees closely with the Cluster Expansion (CE)~\cite{muzyk2011phase} and experimental~\cite{blum2005prediction} results over much of the composition range, whereas the EAM potential predicts a substantially deeper minimum.

\subsubsection{Stable Ordered Structures from Convex-Hull Analysis}

\begin{figure}[htbp]
    \centering
    \includegraphics[width=0.5\textwidth]{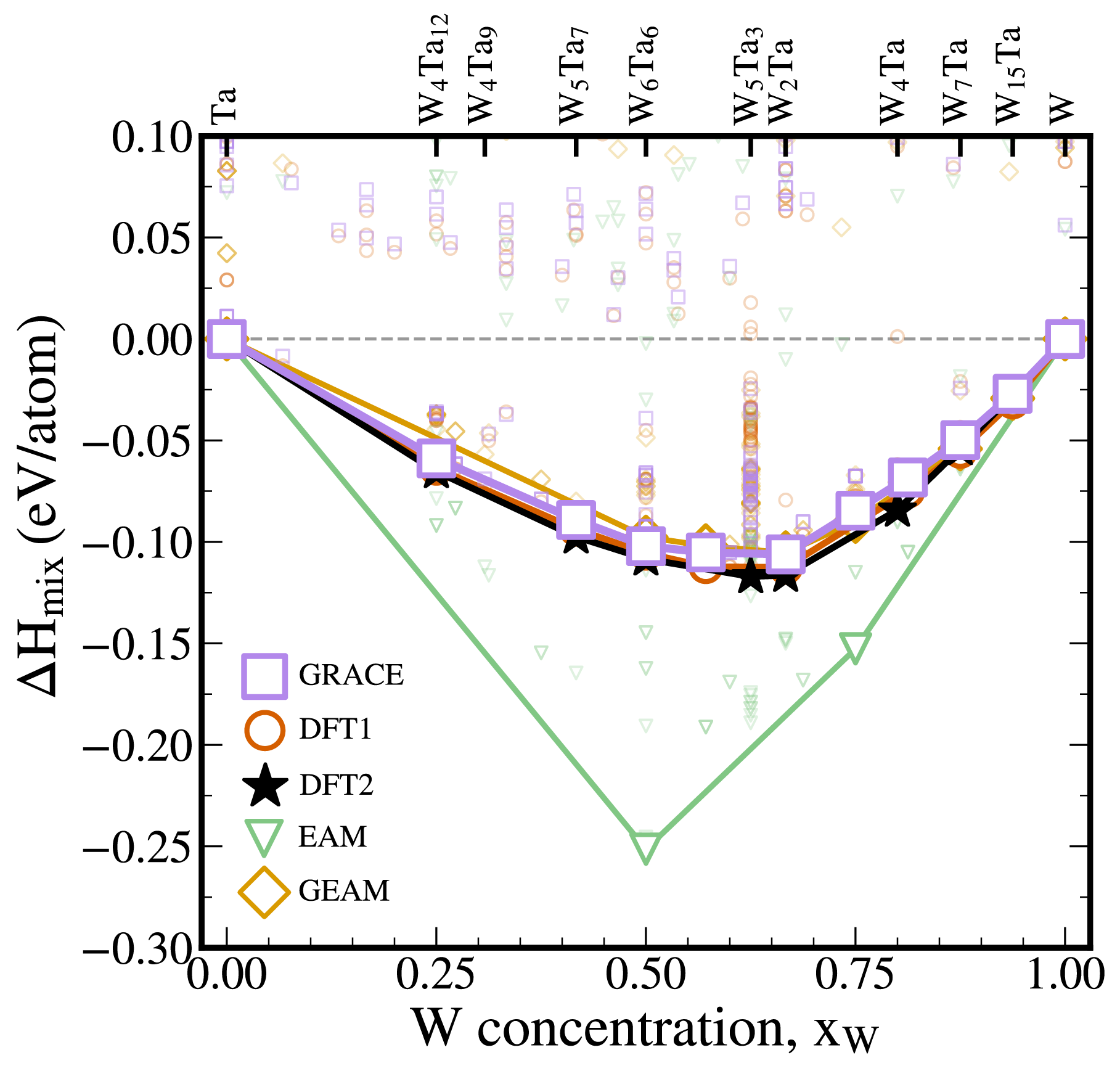}
    \caption{Convex hull of the W--Ta mixing enthalpy, $\Delta H_{\mathrm{mix}}$, as a function of W concentration, $x_{\mathrm{W}}$. Solid lines indicate the corresponding lower convex hulls. Structures on the hull are stable ground states, whereas the smaller, semi-transparent symbols above the hull represent metastable structures. The dashed line marks $\Delta H_{\mathrm{mix}}=0$.}
    \label{fig:wta_convex_hull}
\end{figure}

The thermodynamic stability of ordered W--Ta compounds was evaluated using a convex-hull analysis of the mixing enthalpy, as shown in Fig.~\ref{fig:wta_convex_hull}. Structures lying on the convex hull represent zero-temperature ground states, whereas structures above the hull are unstable with respect to decomposition into neighboring stable phases.

Although the experimental W--Ta phase diagram indicates complete high-temperature solid solubility without stable intermetallic compounds, first-principles studies have reported conflicting predictions for the low-temperature ground states. Calculations~\cite{turchi2001first} predicted ordered $\mathrm{B}_2$ and $\mathrm{DO}_3$ phases, while cluster-expansion studies identified different stable structures, including $\mathrm{C11}_b$~\cite{blum2005prediction} and $\mathrm{W}_{5}\mathrm{Ta}_{3}$~\cite{muzyk2011phase}.

The candidate structures used for the convex-hull construction were taken from the Open Quantum Materials Database (OQMD)~\cite{saal2013materials}, previous DFT calculations for the W--Ta system~\cite{muzyk2011phase}, and randomly generated intermetallic compounds. The calculated convex hull exhibits several stable intermediate compositions between the pure Ta and pure W, with the lowest mixing enthalpies occurring near $\mathrm{W}_{2}\mathrm{Ta}$ compositions. In Fig.~\ref{fig:wta_convex_hull}, the DFT1 data correspond to the present DFT calculations performed for the same set of structures used in the GRACE convex-hull exploration. The DFT2 data are taken from Ref.~\cite{muzyk2011phase}. For direct comparison, the same structural set was also evaluated using the EAM and GEAM potentials to construct the corresponding convex hulls.

GRACE reproduces the shape of the DFT convex hull and predicts stable ordered structures across the composition range, consistent with the favorable alloying tendency of the W--Ta system. Notably, GRACE also provides accurate predictions in the W-rich, low-Ta concentration regime. This regime is particularly important for modeling the microstructural evolution of W--Ta alloys considered for plasma-facing component applications, where small Ta additions can influence defect behavior, phase stability, and irradiation response.

\subsubsection{Melting Temperature}
The W--Ta phase diagram has a lens-shaped form with a narrow solidus--liquidus gap~\cite{turchi2007interface}, making the individual phase boundaries difficult to resolve directly using atomistic simulations. Therefore, the solid--liquid coexistence temperature at each composition was used as a practical estimate of the melting behavior.

Random BCC W--Ta solid solutions were constructed in $6\times6\times50$ supercells with periodic boundary conditions. Each structure was equilibrated near an estimated melting temperature, $T_0$, for $1~\mathrm{ns}$ under zero-pressure NPT conditions. An interior slab was then melted at $5000~\mathrm{K}$ for $100~\mathrm{ps}$ while the surrounding solid was fixed. The full system was subsequently evolved for $10~\mathrm{ns}$ in the NPH ensemble with pressure control applied along the $z$ direction. The stabilized average temperature was taken as the melting temperature; calculations that fully melted or crystallized were repeated using an updated $T_0$.

\begin{figure}[H]
\centering
\includegraphics[width=0.65\textwidth]{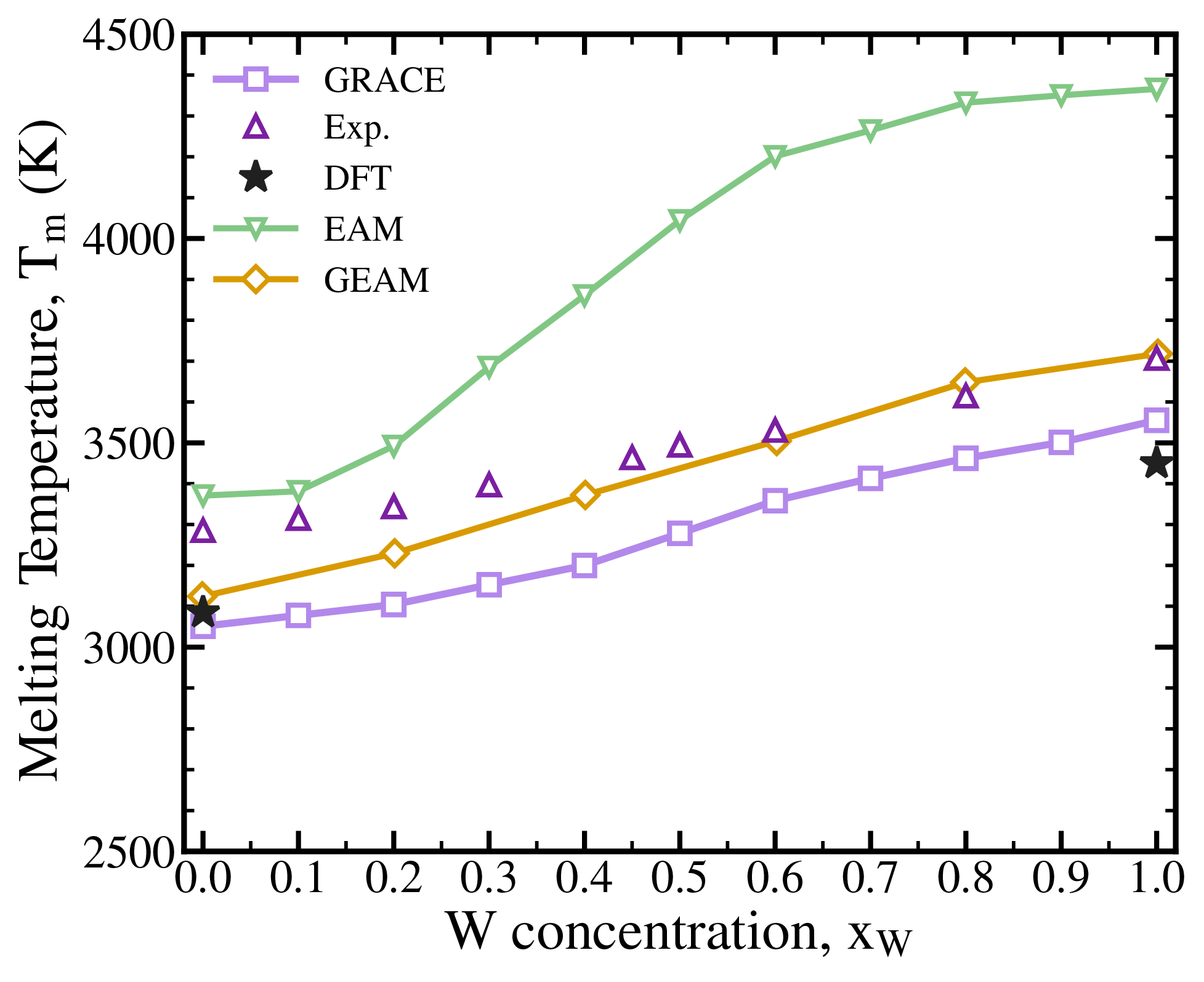}
\caption{
Melting temperature of the W--Ta binary system as a function of W concentration, $x_{\mathrm{W}}$.}
\label{fig:melting_temp}
\end{figure}

As shown in Fig.~\ref{fig:melting_temp}, GRACE predicts a monotonic increase in melting temperature with increasing W concentration. Although it underestimates the experimental values~\cite{rudy1969compendium} by less than $10\%$, its predictions are closer to the DFT melting temperatures of the pure elements~\cite{taioli2007melting,wang2011melting} and reproduce the overall composition dependence. GEAM shows closer agreement with experiment~\cite{rudy1969compendium}, whereas EAM substantially overestimates the melting temperature, particularly in the W-rich region.

\section{GRACE Application: Effects of Ta Segregation on Grain-Boundary Pinning in W}

\subsection{Motivation}

During high-temperature exposure and mechanical loading, W undergoes recovery, recrystallization, and grain growth, gradually losing the strengthened microstructure produced by deformation. GB chemistry can strongly affect this evolution because GBs control the growth of recrystallized grains. Experiments on nanocrystalline W alloys have shown that solute additions can suppress abnormal grain growth and improve microstructural stability, although the degree of stabilization depends strongly on how the solute is distributed along the GBs~\cite{donaldson2018solute}. In W--Ta alloys containing approximately 1--5\% Ta, Ta additions have been reported to increase the recrystallization temperature, limit grain coarsening, and improve strength retention during high-temperature annealing~\cite{nogami2022tungsten}.

GBs also play important roles in intergranular fracture, irradiation-defect absorption, and high-temperature deformation in W-based alloys. Under applied shear or other driving forces, GBs can respond through coupled migration~\cite{cahn2006coupling}, sliding~\cite{ivanov2008dynamics}, or intermittent stick--slip motion~\cite{mishin2007stick}. During stick--slip motion, the boundary remains temporarily pinned while elastic strain builds up, followed by a rapid displacement that releases the stored stress. This displacement may occur through the motion of disconnections, GB dislocations, twinning partials, or coordinated atomic rearrangements. The dominant mechanism depends strongly on the GB structure. For example, stick--slip has been linked to the nucleation of twinning partials in Cu~\cite{hu2009stick}, the climb of GB dislocations in SiC~\cite{bringuier2015grain}, and the interruption of ordered atomic rearrangements near coincidence-site-lattice atoms in Ni~\cite{bair2019antithermal}. GB chemistry can further change whether a boundary migrates, slides, or remains pinned by altering the energy required for each process~\cite{xing2024transition}. Solute segregation has been shown to locally pin Ni--Cu GBs~\cite{verma2024solute} and to hinder migration while promoting sliding in refractory MoNbTa alloys~\cite{tan2026competition}. These findings suggest that alloying can affect not only the overall GB mobility but also where and how strongly a boundary becomes locally pinned.

Despite these observations, the atomic-scale behavior of Ta at W GBs remains poorly understood. First-principles calculations supported by transmission electron microscopy showed that Ta has little tendency to segregate to the investigated low-angle W GBs, but has a modest preference for selected sites within high-angle GB cores. This result demonstrates that Ta segregation depends strongly on GB character and local atomic structure~\cite{almotasem2021factors}. More recent experimental and atomistic studies of sputtered W--10 at.\% Ta found that Ta remained primarily distributed as a solid solution rather than forming the solute-rich clusters observed for other alloying elements~\cite{almotasem2025exploring}. The influence of Ta on recrystallization, grain growth, and GB deformation in W therefore cannot be described solely by the nominal alloy concentration. It must also account for GB character, temperature, finite-composition effects, and the occupation of specific GB sites. In particular, it remains unclear how Ta segregation changes local pinning and stick--slip de-pinning in W. The present work therefore investigates finite-temperature Ta segregation at a $\Sigma5(120)$ symmetric-tilt W GB and determines how the resulting site-specific solute distributions affect GB migration and de-pinning.

\subsection{Simulation Setup}

\begin{figure}[H]
\centering
\includegraphics[width=\textwidth]{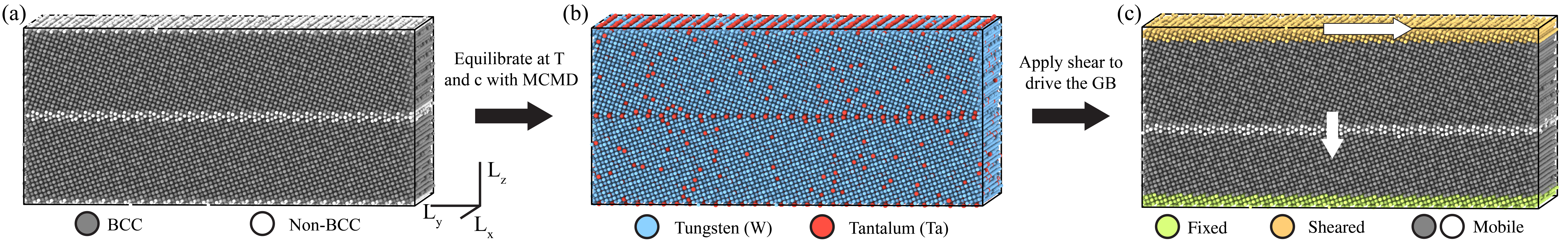}
\caption{Simulation workflow used to construct, equilibrate, and shear the W--Ta $\Sigma5(120)$ symmetric-tilt GB. (a) Minimum-energy pure-W GB configuration obtained using the GRIP method. BCC and non-BCC W atoms are shown in gray and white, respectively. (b) Ta atoms are randomly introduced at the prescribed nominal composition, followed by chemical and structural equilibration using hybrid MC/MD simulations. W and Ta atoms are shown in blue and red, respectively. (c) Shear-loading configuration used to drive GB migration. The lower terminal slab is fixed, the upper slab is translated parallel to the GB plane at $1~\mathrm{m/s}$, and the atoms in the intervening region remain fully mobile and equilibrated at the target temperature.}
\label{fig:simulation_setup}
\end{figure}

A $\Sigma5(120)$ symmetric-tilt GB was generated using the GRand canonical Interface Predictor (GRIP) method~\cite{chen_2024_grip} with an embedded-atom method (EAM) potential~\cite{zhou2004misfit}. Multiple microscopic GB configurations were sampled by varying the interfacial atomic density and the relative translation between the adjoining grains. The configuration with the lowest GB energy was selected for all subsequent simulations, as illustrated in Fig.~\ref{fig:simulation_setup}(a).

The minimum-energy GB structure was initially constructed in pure W and subsequently populated with randomly distributed Ta atoms to obtain nominal compositions of 0, 1, 5, 10, 15, 20~at.\% Ta. Chemical and structural equilibration was performed over the temperature range 500--3000~K using hybrid Monte Carlo/Molecular Dynamics (MC/MD) simulations~\cite{sadigh2010scalable} with the newly developed W--Ta GRACE potential, see Fig.~\ref{fig:simulation_setup}(b). The systems were maintained at zero pressure using an anisotropic isothermal--isobaric ensemble and integrated with a timestep of 1~fs under periodic boundary conditions in all directions. Composition-conserving W--Ta identity-swap attempts were performed every 100 MD steps while maintaining the prescribed global Ta concentration. Each system was initially equilibrated for $2{,}000{,}000$ MD steps, corresponding to 2~ns. The equilibrated structure was then replicated to produce a simulation cell with approximate dimensions of $15\times25\times10~\mathrm{nm^3}$ and equilibrated for an additional 2~ns at the same temperature and composition.

Following equilibration, periodic boundary conditions were retained along the two directions parallel to the GB plane, whereas the boundary condition along the GB-normal direction was changed to non-periodic. Two terminal slabs, each approximately 1~nm thick, were defined at the upper and lower ends of the simulation cell. The lower slab was held fixed, while the upper slab was translated rigidly parallel to the GB plane at a constant velocity of $1~\mathrm{m/s}$, as shown in Fig.~\ref{fig:simulation_setup}(c). Atoms between the terminal slabs remained fully mobile and were maintained at the target temperature in the canonical ensemble. This loading configuration imposed shear across the bicrystal and drove shear-coupled GB migration.

\subsection{Quantification Metrics}
In this study, the influence of Ta on the initial pinning of the W GB was characterized using both global and atomic row-resolved segregation metrics. Here, an atomic row refers to a line of atoms along the [100] crystallographic direction. The average Ta concentration within the GB region, ($c^{\mathrm{GB}}_{\mathrm{Ta}}$), was monitored, and the segregation excess was defined as $\Gamma_{\mathrm{seg}}=c^{\mathrm{GB}}_{\mathrm{Ta}}-c^{\mathrm{bulk}}_{\mathrm{Ta}}$, where ($c^{\mathrm{bulk}}_{\mathrm{Ta}}$) is the Ta concentration in the surrounding bulk solid solution. Concentration profiles normal to the GB were used to determine the spatial extent of Ta enrichment as functions of temperature and nominal composition. Because the GB contains crystallographically distinct atomic rows, the Ta occupancy of each row was also evaluated separately. This row-resolved analysis captures the highly nonuniform segregation state, including the preferential saturation of specific rows, which cannot be represented adequately by the average GB concentration alone. 

The imposed grain translation was calculated from the simulation timestep and the prescribed shear velocity as ($d=v_{\mathrm{shear}}N\Delta t$), where ($N$) is the timestep index and ($\Delta t$) is the MD timestep. The corresponding shear stress, ($\sigma_{yz}$), was obtained from the global atomic virial stress. The first maximum in ($\sigma_{yz}$) preceding the onset of GB migration was defined as the \emph{de-pinning stress}. This quantity represents the initial stress required to overcome the resistance produced by the Ta segregation at the boundary and initiate motion of the GB away from its pinned configuration.

\subsection{Simulation Results}

\subsubsection{Ta Segregation}

Ta segregation to the $\Sigma5(120)$ tilt GB was examined over temperatures ranging from 500 to 3000~K and nominal Ta concentrations ranging
from 1 to 20~at.\%. The equilibrium segregation excess was calculated as $\Gamma_{\mathrm{seg}}=c_{\mathrm{Ta}}^{\mathrm{GB}}-c_{\mathrm{Ta}}^{\mathrm{bulk}}$, where $c_{\mathrm{Ta}}^{\mathrm{GB}}$ and $c_{\mathrm{Ta}}^{\mathrm{bulk}}$ are the equilibrium Ta concentrations within
the GB region, defined with 1.5 nm width, and the adjoining bulk grains, respectively. A positive value of $\Gamma_{\mathrm{seg}}$ therefore indicates preferential Ta enrichment at the GB.

\begin{figure}[H]
    \centering
    \includegraphics[width=0.70\textwidth]{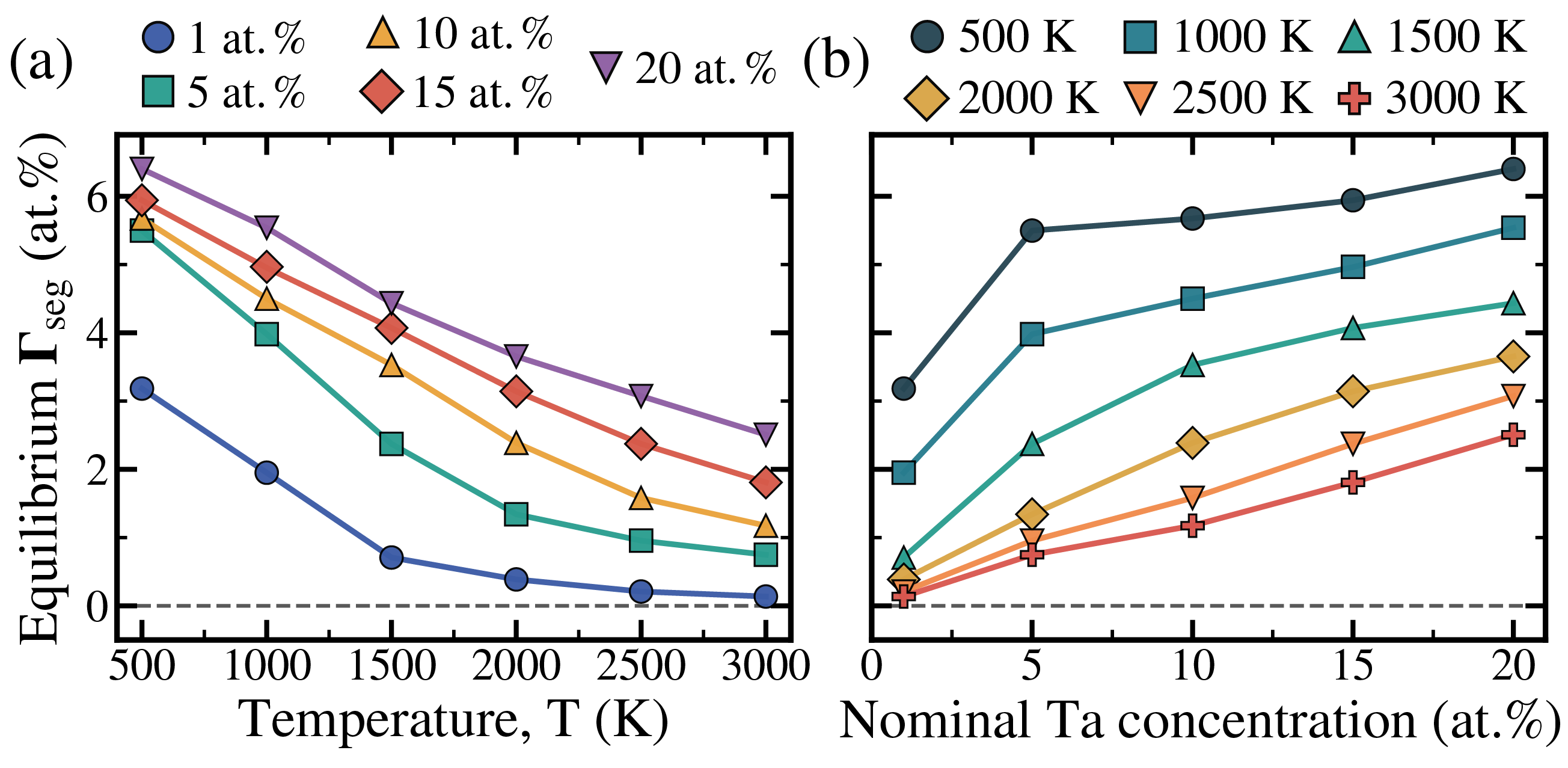}
    \caption{
        Equilibrium Ta segregation at the $\Sigma5(120)$ tilt GB. (a) Equilibrium segregation excess, $\Gamma_{\mathrm{seg}} = c_{\mathrm{Ta}}^{\mathrm{GB}}-c_{\mathrm{Ta}}^{\mathrm{bulk}}$,as a function of temperature for different nominal Ta concentrations. (b) Equilibrium segregation excess as a function of nominal Ta concentration at temperatures between 500 and 3000~K. The horizontal dashed line denotes the condition of no preferential segregation, $\Gamma_{\mathrm{seg}}=0$.
    }
    \label{fig:Ta_seg_S5}
\end{figure}

Figure~\ref{fig:Ta_seg_S5}(a) shows that the segregation excess decreases monotonically with increasing temperature for all nominal Ta concentrations. This behavior is expected and indicates that thermal fluctuations progressively weaken the preferential occupation of GB sites by Ta. The temperature dependence is particularly pronounced at low nominal concentration: for the 1~at.\% system, $\Gamma_{\mathrm{seg}}$ approaches zero at the highest temperatures considered. In contrast, systems with higher nominal Ta concentrations retain a measurable segregation excess over the entire temperature range. At a fixed temperature, $\Gamma_{\mathrm{seg}}$ increases with nominal Ta concentration, as shown in Fig.~\ref{fig:Ta_seg_S5}(b). Thus, although elevated temperatures reduce
segregation, increasing the available Ta concentration promotes continued enrichment of the GB.

Although $\Gamma_{\mathrm{seg}}$ quantifies the degree of partitioning between the GB and bulk regions, the present study focuses primarily on the first GB \emph{de-pinning} event. The onset of de-pinning is governed most directly by the local chemical and structural state of the boundary, particularly the number and
spatial distribution of Ta atoms occupying the GB sites. The bulk composition acts as the Ta reservoir, but it has a less direct influence on the initial
de-pinning event than the instantaneous Ta concentration within the GB.
Consequently, the following analysis emphasizes $c_{\mathrm{Ta}}^{\mathrm{GB}}$ rather than the segregation excess alone.

\begin{figure}[H]
    \centering
    \includegraphics[width=\textwidth]{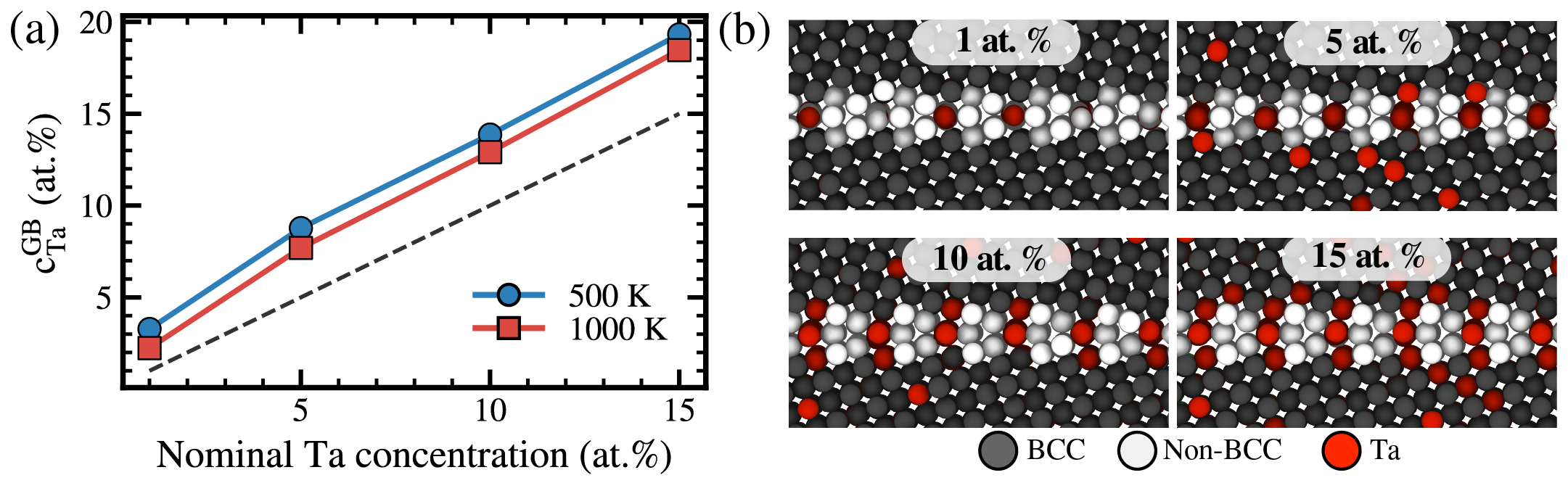}
    \caption{
        Ta enrichment and spatial distribution within the $\Sigma5(120)$ GB. (a) Equilibrium GB Ta concentration, $c_{\mathrm{Ta}}^{\mathrm{GB}}$, as a function of nominal Ta concentration at 500 and 1000~K. The dashed line represents $c_{\mathrm{Ta}}^{\mathrm{GB}}=c_{\mathrm{Ta}}^{0}$ and therefore corresponds to the absence of GB enrichment. (b) Representative atomistic configurations illustrating the distribution of Ta atoms within the GB at nominal concentrations of 1, 5, 10, and 15~at.\%. BCC atoms, non-BCC GB atoms, and Ta atoms are shown in gray, white, and red, respectively.
    }
    \label{fig:Ta_concs_S5}
\end{figure}

As shown in Fig.~\ref{fig:Ta_concs_S5}(a), the equilibrium GB concentration increases approximately linearly with nominal Ta concentration at both 500 and 1000~K. In every case, $c_{\mathrm{Ta}}^{\mathrm{GB}}$ lies above the no-segregation reference dashed line, confirming that the GB is enriched in Ta relative to the overall alloy composition. The enrichment is slightly greater at 500~K than at 1000~K, consistent with the temperature dependence of $\Gamma_{\mathrm{seg}}$ in Fig.~\ref{fig:Ta_seg_S5}. However, the atomistic configurations in Fig.~\ref{fig:Ta_concs_S5}(b) demonstrate that the segregated Ta atoms are not distributed uniformly throughout the boundary. Instead, Ta preferentially occupies a subset of the available GB sites, producing a spatially heterogeneous solute distribution.

\begin{figure}[H]
    \centering
    \includegraphics[width=\textwidth]{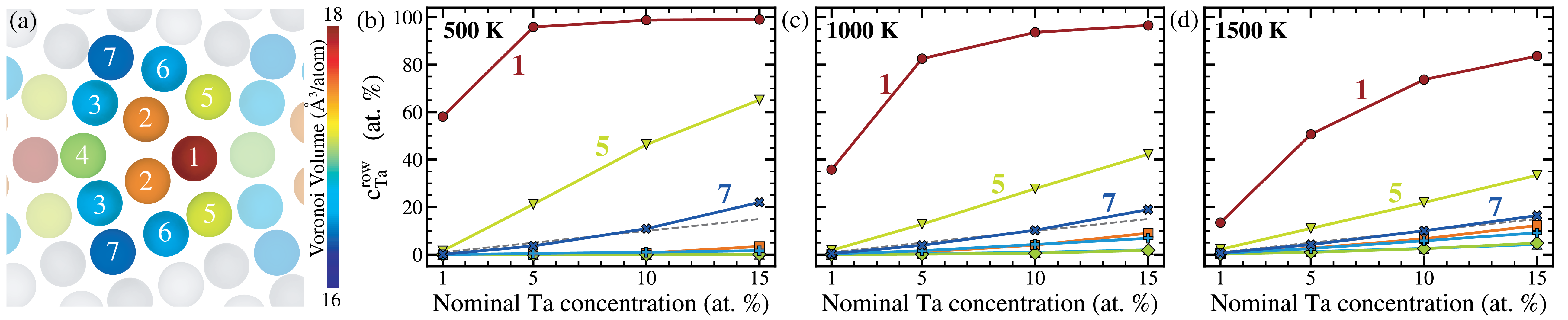}
    \caption{Row-resolved Ta segregation within the $\Sigma5(120)$ tilt GB. (a) Atomistic representation of the GB structural unit showing the seven distinct atomic rows. Each row is colored according to its atomic Voronoi volume, with blue and red corresponding to smaller and larger volumes, respectively. The color scale spans approximately 16--18~\AA$^{3}$/atom. (b--d) Equilibrium Ta concentration within each row as a function of nominal Ta concentration at (b) 500~K, (c) 1000~K, and (d) 1500~K. The curve colors correspond to the mean Voronoi-volume colors assigned in panel~(a), and the dashed line denotes equality between the row and nominal Ta concentrations.}
    \label{fig:strings_seg_S5}
\end{figure}

To resolve the spatial heterogeneity of Ta segregation, the GB core was decomposed into the seven structurally distinct atomic rows shown in Fig.~\ref{fig:strings_seg_S5}(a). For each row, the mean Voronoi volume was calculated by averaging the Voronoi volumes of all atoms assigned to that row using OVITO’s Voronoi analysis modifier~\cite{ovito}. The rows in Fig.~\ref{fig:strings_seg_S5}(a) are colored according to their  Voronoi volume, with blue and red representing relatively small and large local atomic volumes, respectively. The same color assignment is used for the corresponding row-resolved segregation curves in Figs.~\ref{fig:strings_seg_S5}(b)--\ref{fig:strings_seg_S5}(d).

The results reveal a strongly site-selective and sequential segregation process. Row~1, which has the largest Voronoi volume among the identified GB rows, exhibits the strongest Ta enrichment at all temperatures. At 500~K, its Ta concentration is already approximately 60~at.\% at a nominal concentration of 1~at.\%, and it approaches complete occupation by approximately 5~at.\% nominal Ta. Once this high-affinity row approaches saturation, additional Ta is increasingly accommodated by Row~5, which also possesses a relatively large Voronoi volume. Row~5 consequently shows a continuous increase in Ta concentration with increasing nominal alloy composition.

Row~7, which has a smaller mean Voronoi volume, displays substantially weaker enrichment than rows~1 and 5, although its occupation becomes appreciable at higher nominal Ta concentrations. Most other rows exhibit Ta concentrations near or below the nominal alloy concentration, particularly at 500~K. These observations suggest that Ta preferentially occupies GB rows with larger local atomic volumes, where the additional space may accommodate the solute with less local distortion.

The preference for high-volume rows persists at 1000 and 1500~K, but the magnitude of the site-specific enrichment decreases as temperature increases. At 1000~K, row~1 remains the dominant segregation site and approaches saturation at high nominal Ta concentration, while occupation of row~5 increases more gradually. At 1500~K, enrichment of row~1 remains pronounced, but complete saturation is not reached within the investigated composition range. The concentration differences among the lower-affinity rows also become less pronounced with increasing temperature because thermal fluctuations promote a broader distribution of Ta among the available GB sites.

Although Ta preferentially occupies rows with relatively large Voronoi volumes, the correlation is not strictly monotonic. The segregation preference
also depends on the local coordination, distortion, and chemical environment. Moreover, complete saturation of one row may redistribute the GB excess volume and modify the local Voronoi volumes, thereby favoring sites that initially exhibited smaller Voronoi volumes in the pure-W GB. Thus, Voronoi volume is an important but composition-dependent descriptor of site-specific segregation.

\subsubsection{Ta-Induced Pinning of W GB}

\begin{figure}[H]
    \centering
    \includegraphics[width=\textwidth]
    {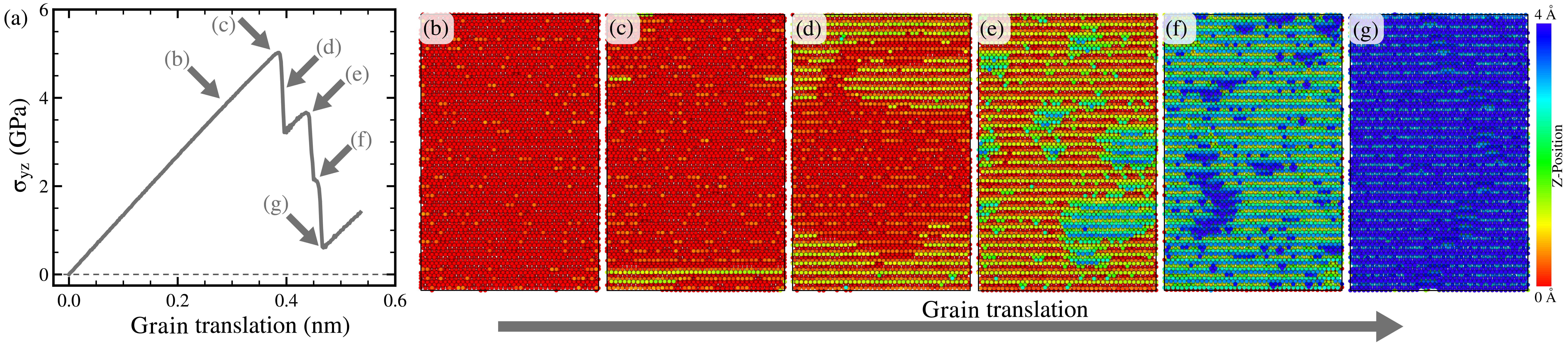}
    \caption{
    Disconnection-mediated de-pinning of the GB containing 5~at.\% Ta at
    500~K. (a) Applied shear stress, $\sigma_{yz}$, as a function of grain
    translation. The first stress maximum, approximately 5~GPa, marks the
    onset of GB de-pinning. (b--g) Bottom-view atomic configurations at the
    locations indicated in panel (a), with atoms colored by their
    $z$-position. The sequence shows localized disconnection nucleation,
    propagation along the GB, and the resulting two-stage stress release
    preceding nearly complete GB detachment.
    }
    \label{fig:Disconnection_Nucleation}
\end{figure}

The atomistic mechanism associated with the first GB de-pinning event is illustrated in Fig.~\ref{fig:Disconnection_Nucleation} for the system containing 5~at.\% Ta at 500~K. The applied shear stress, $\sigma_{yz}$, initially increases approximately linearly with grain translation and reaches a maximum of approximately 5~GPa. The atomic snapshots corresponding to selected points along the stress--translation curve show that GB migration does not occur through a spatially uniform translation of the entire interface. Instead, a localized disconnection first nucleates along the GB and subsequently propagates along the interface. The resulting motion produces two pronounced stress-release stages. The first stress drop corresponds to initial propagation of the disconnection and migration of the GB by $\approx 2~\AA$, whereas the subsequent drop accompanies further propagation and nearly complete detachment of the boundary from its initial configuration. Thus, the first maximum in $\sigma_{yz}$ is identified as the GB de-pinning stress associated with the nucleation and initial propagation of a disconnection.

\begin{figure}[H]
    \centering
    \includegraphics[width=\textwidth]
    {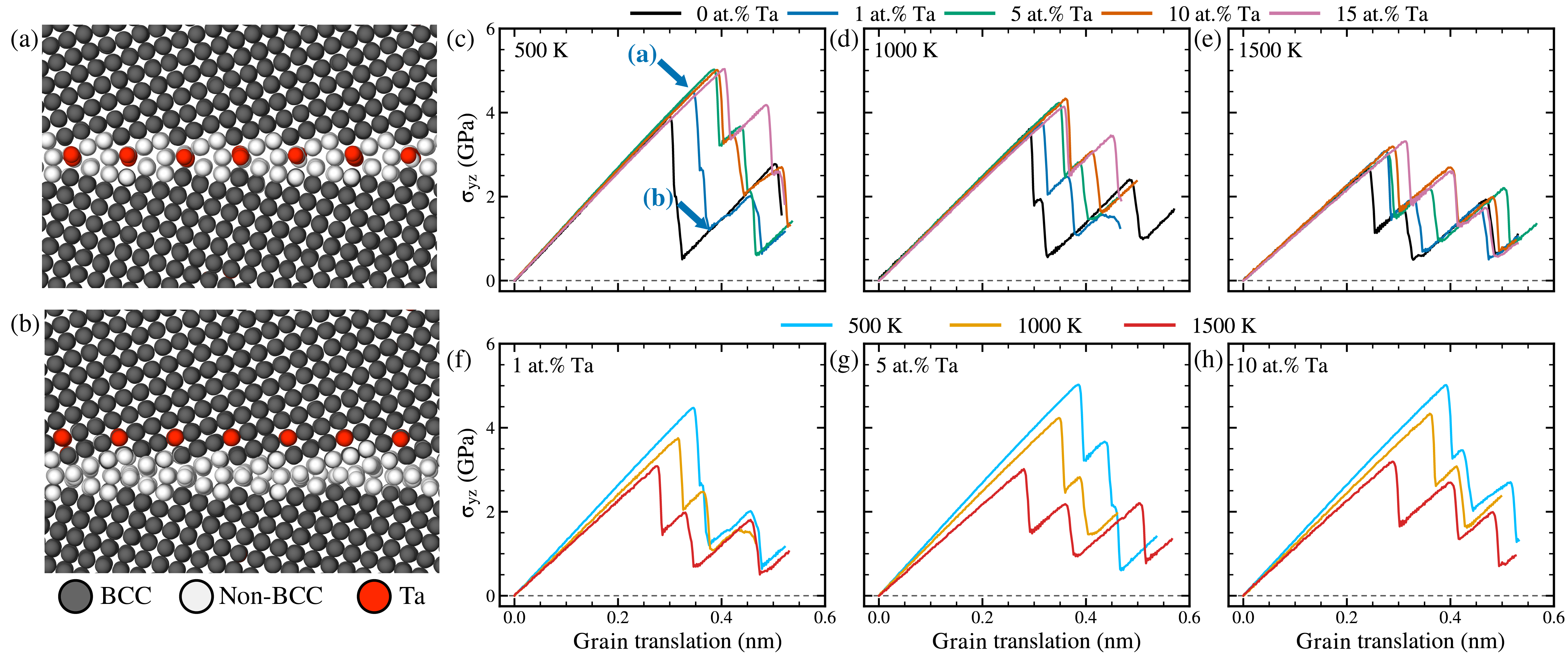}
    \caption{
    Effects of Ta concentration and temperature on GB de-pinning. (a,b) Atomic configurations of the 1~at.\% Ta system at 500~K immediately before and after the first de-pinning event, respectively. BCC atoms are shown in gray, non-BCC GB atoms in white, and Ta atoms in red. (c--e) Applied shear stress, $\sigma_{yz}$, as a function of grain translation for different Ta concentrations at 500, 1000, and 1500~K, respectively. (f--h) Stress--translation curves grouped by temperature for systems containing 1, 5, and 10~at.\% Ta, respectively.}
    \label{fig:Stress_curves}
\end{figure}

The influence of Ta concentration and temperature on the de-pinning response is summarized in Fig.~\ref{fig:Stress_curves}. The configurations in Figs.~\ref{fig:Stress_curves}(a) and \ref{fig:Stress_curves}(b) show the solute-decorated GB immediately before and after the first de-pinning event for the 1~at.\% Ta system at 500~K. The non-BCC atoms identify the migrating GB region, whereas the Ta atoms remain essentially stationary as the GB detaches and migrates away from the segregated solute sites. Their interaction with the GB therefore increases the applied stress required to initiate migration.

At 500~K, even the addition of 1~at.\% Ta produces a noticeable increase in the first de-pinning stress relative to pure W. The de-pinning stress increases further between 1 and 5~at.\% Ta, but changes only weakly between approximately 5 and 15~at.\% Ta. This plateau indicates that the dominant segregation configurations controlling the first de-pinning event become nearly saturated. Once these high-impact sites or rows are occupied, additional Ta atoms are distributed among less influential GB sites and produce only a limited additional increase in the de-pinning stress. A similar, although weaker, concentration dependence is observed at 1000~K, whereas the effect of Ta concentration becomes substantially less pronounced at 1500~K.

Figures~\ref{fig:Stress_curves}(f--h) reorganize the same results by Ta concentration and demonstrate the strong influence of temperature. For each examined composition, the first peak in $\sigma_{yz}$ decreases as the temperature increases from 500 to 1500~K. This reduction is consistent with thermally assisted disconnection nucleation and propagation: increasing temperature enhances atomic fluctuations and reduces the mechanical stress required for the GB to escape from the segregated configuration. Ta segregation consequently produces its strongest pinning effect at 500~K, while thermal activation progressively weakens this effect at higher temperatures.

\begin{figure}[H]
    \centering
    \includegraphics[width=\textwidth]
    {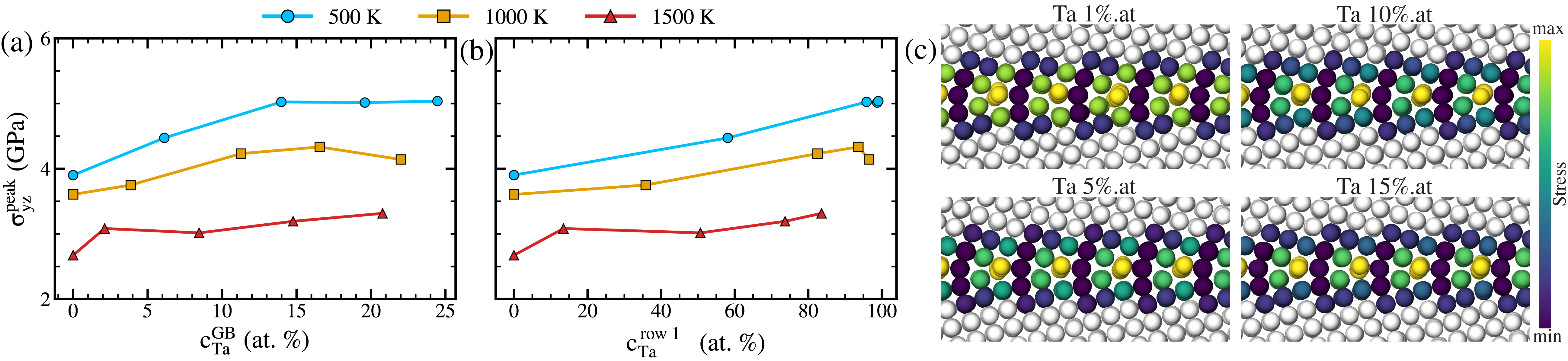}
    \caption{Relationship between Ta segregation, local stress redistribution, and GB de-pinning strength. The first de-pinning stress is plotted as a function of (a) the overall GB Ta concentration, $c_{\mathrm{Ta}}^{\mathrm{GB}}$, and (b) the Ta concentration within row~1, $c_{\mathrm{Ta}}^{\mathrm{row\,1}}$. The increase and subsequent saturation of the peak stress correlate with progressive occupation of row~1. (c) Atomic configurations for nominal Ta concentrations of 1, 5, 10, and 15~at.\%, with tracked GB-row atoms colored by the row-averaged normalized difference in $\sigma_{xx}$ relative to the pure-W peak configuration. The color scale ranges from approximately $-2$ to $20$. The largest relative stress changes are localized on the segregation-rich row~1.}
    \label{fig:Peak_Stress}
\end{figure}

To distinguish the influence of the total GB composition from that of solute occupation at specific atomic rows, the first de-pinning stress was plotted against both the overall GB Ta concentration, $c_{\mathrm{Ta}}^{\mathrm{GB}}$, and the Ta concentration within row~1, $c_{\mathrm{Ta}}^{\mathrm{row\,1}}$, as shown in Fig.~\ref{fig:Peak_Stress}. At 500~K, the de-pinning stress initially increases with the overall GB Ta concentration and subsequently approaches a plateau. A corresponding relationship is obtained when the stress is plotted against the concentration of row~1. In particular, the de-pinning stress increases as row~1 becomes progressively occupied and changes only weakly once its Ta concentration approaches saturation. The correlation with the occupation of this individual row indicates that the first de-pinning event is controlled more directly by the local solute configuration than by the average GB concentration alone.

The local stress redistribution associated with this segregation was quantified by comparing the row-averaged $\sigma_{xx}$ component at the first stress peak of each Ta-containing system with the corresponding row-averaged value at the first peak of the pure-W reference system.
For each physical row $s$, the normalized difference was calculated as $\Delta\overline{\sigma}_{xx}^{(s)} = (\overline{\sigma}_{xx,\mathrm{Ta}}^{(s)}-\overline{\sigma}_{xx,0}^{(s)})/(\overline{\sigma}_{xx,0}^{(s)})$, where $\overline{\sigma}_{xx,\mathrm{Ta}}^{(s)}$ is the mean value of $\sigma_{xx}$ over all atoms assigned to row $s$ in the Ta-containing peak configuration, and $\overline{\sigma}_{xx,0}^{(s)}$ is the corresponding mean for the pure-W peak configuration. The atoms belonging to each row were identified in the initial configuration and used for subsequent tracking, ensuring that the same physical groups of atoms were compared between configurations. Each atom belonging to a given row was assigned the same value of $\Delta\overline{\sigma}_{xx}^{(s)}$ for visualization. The mapped quantity is normalized and dimensionless. The color range in Fig.~\ref{fig:Peak_Stress}(c) therefore extends from approximately $-2$ to $20$. The $\sigma_{xx}$ component was selected for this visualization because it provides the clearest spatial contrast among the different GB rows. The maps show that the largest relative row-averaged stress changes are concentrated on the segregation-rich row~1. This localization, together with the correlation between $c_{\mathrm{Ta}}^{\mathrm{row\,1}}$ and the first stress peak, supports the conclusion that occupation of this specific GB environment strongly controls the pinning strength and the initiation of disconnection-mediated de-pinning.

The preceding results establish a strong correlation between the occupation of row~1 and the first GB de-pinning stress. To test this relationship more directly and separate the effect of individual GB environments from the complex equilibrium segregation distribution, controlled row-occupation simulations were performed. Starting from the pure-W $\Sigma5(120)$ GB, all W atoms belonging to one selected GB row were artificially replaced with Ta atoms, while the remaining GB rows and bulk regions were retained as pure W. Each resulting configuration was relaxed at 500~K and zero pressure using the anisotropic NPT ensemble. The relaxed structures were then subjected to the same shear-loading setup used for the equilibrated W--Ta systems.

\begin{figure}[H]
    \centering
    \includegraphics[width=\textwidth]{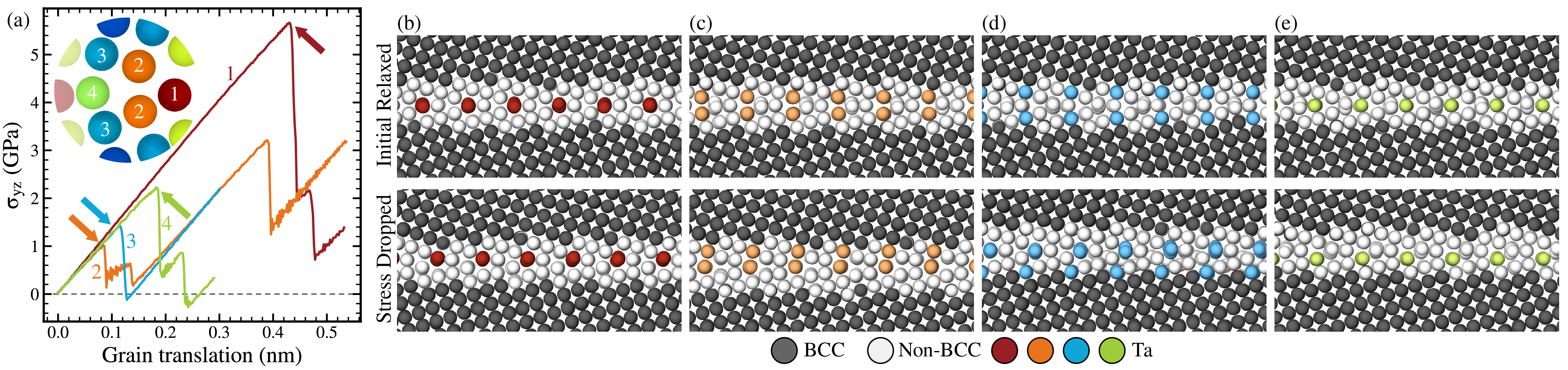}
    \caption{
    Controlled row Ta occupation in the $\Sigma5(120)$ W GB at 500~K. Each selected W row was replaced entirely by Ta, relaxed at zero pressure in the NPT ensemble, and subjected to shear loading. (a) $\sigma_{yz}$ as a function of grain translation for Ta occupation of rows~1--4; arrows mark the first de-pinning stress drop. Occupation of row~1 produces the highest peak stress, approximately $5.7$~GPa. (b--e) Relaxed configurations (top) and configurations after the first stress drop (bottom) for rows~1--4. Bulk W, GB atoms, and Ta atoms are shown in dark gray, white, and the corresponding row colors, respectively. The results identify row~1 as the dominant site controlling initial GB pinning.
    }
    \label{fig:controlled_string_pinning}
\end{figure}

Figure~\ref{fig:controlled_string_pinning}(a) compares the resulting $\sigma_{yz}$--grain-translation curves for complete Ta occupation of rows~1--4. These configurations are controlled mechanistic tests rather than equilibrium segregation states; therefore, the comparison evaluates the effect of occupying each structural row and does not represent a comparison at identical global Ta concentration. The arrows identify the first pronounced stress drop in each simulation, which is taken as the initial GB de-pinning event. Ta occupation of row~1 produces a markedly different response from occupation of the other rows. The shear stress continues to increase to approximately $5.7$~GPa at a grain translation of approximately $0.43$~nm before undergoing a sharp decrease. In contrast, occupation of rows~2, 3, and 4 produces substantially earlier de-pinning at first peak stresses of approximately $1.1$, $1.4$, and $2.2$~GPa, respectively.

The atomic configurations before loading and immediately after the first stress drop are shown in Figs.~\ref{fig:controlled_string_pinning}(b)--\ref{fig:controlled_string_pinning}(e). In each case, the stress drop is accompanied by displacement of the GB region relative to its initially relaxed configuration. The smaller and earlier stress drops obtained for rows~2--4 indicate that Ta occupation of these environments provides only limited resistance to the initiation of GB motion. Subsequent increases in stress correspond to continued loading of the newly migrated GB configuration and should therefore be distinguished from the first de-pinning event.

The substantially greater stress and grain translation required to de-pin the GB from Ta-occupied row~1 provide direct evidence that the different GB rows do not contribute equally to solute pinning. Instead, the pinning strength is governed strongly by the crystallographic and local atomic environment occupied by Ta. This controlled comparison supports the interpretation of the equilibrium-segregation simulations: the increase and eventual plateau in the first de-pinning stress arise primarily from the progressive occupation and saturation of row~1, whereas additional Ta distributed among lower-impact rows produces a comparatively small increase in the initial resistance to GB migration.

\section{Conclusion}

We developed a transferable GRACE machine-learning interatomic potential for the W–Ta binary system using a large and diverse DFT dataset generated primarily through feature-space information-entropy maximization. Rather than relying on the manual enumeration of anticipated structures, defects, and deformation modes, this approach automatically explores configuration space by maximizing the diversity of atomically relevant features. It therefore enables broad training datasets to be constructed without iterative human selection of configurations or prior knowledge of every downstream application. Supplementing this automatically generated dataset with ordered alloy structures produced a potential that achieves substantially lower energy and force errors than the available empirical W–Ta potentials on the present diverse test set.

Extensive validation demonstrates that the breadth introduced during data generation translates into broad physical transferability. The potential provides a consistent description of equilibrium structures, elastic and vibrational properties, high-pressure response, surfaces, point defects, dislocation energetics, transformation pathways, alloy thermodynamics, and melting behavior.  Taken together, these tests show that entropy-maximized sampling can produce a model capable of describing physical phenomena that were not individually targeted during dataset construction. The resulting potential provides a robust foundation for simulations of the coupled mechanical, defect, and microstructural processes relevant to W–Ta alloys in fusion environments.

As a demonstration of these capabilities, the potential was applied to Ta segregation and shear-coupled migration of a $\Sigma5(120)$ W GB. Ta was found to segregate preferentially to specific atomic rows within the boundary, and occupation of one high-affinity environment strongly increased the stress required for disconnection-mediated de-pinning. The de-pinning stress plateaued as this dominant site became saturated and decreased with increasing temperature. Controlled row-occupation simulations confirmed that pinning depends more strongly on the local atomic environment occupied by Ta than on nominal composition or average GB concentration alone.

\bibliographystyle{unsrt}
\bibliography{ref}
\end{document}